\documentclass{aa} 

\usepackage{graphicx}

\usepackage{txfonts}
\begin{document}

   \title{Low-frequency radio spectra of submillimetre galaxies in the Lockman Hole}

   \subtitle{}

   \author{J. Ramasawmy
          \inst{1}
          \and
          J. E. Geach
          \inst{1}
          \and
          M. J. Hardcastle
          \inst{1}
          \and
          P. N. Best
          \inst{2}
          \and
          M. Bonato
          \inst{3,4,5}
          \and
          M. Bondi
          \inst{3}
          \and
          G. Calistro Rivera
          \inst{6}
          \and
          R. K Cochrane
          \inst{7}
          \and
          J. E. Conway
          \inst{8}
          \and
          K. Coppin
          \inst{1}
          \and
          K.J. Duncan
          \inst{9, 2}
          \and
          J.S. Dunlop
          \inst{2}
          \and
          M. Franco
          \inst{1}
          \and
          C. Garc\'ia-Vergara
          \inst{9}
          \and
          M. J. Jarvis
          \inst{10, 11}
          \and
          R. Kondapally
          \inst{2}
          \and
          I. McCheyne
          \inst{12}
          \and
          I. Prandoni
          \inst{3}
          \and
          H. J. A. R\"ottgering
          \inst{9}
          \and
          D. J. B. Smith
          \inst{1}
          \and
          C. Tasse
          \inst{13,14,15}
          \and
          L. Wang
          \inst{16,17}
          }

   \institute{Centre for Astrophysics Research, Department of Physics, Astronomy and Mathematics, University of Hertfordshire, College Lane, Hatfield AL10 9AB, UK \\
              \email{j.ramasawmy@herts.ac.uk}
    \and
    SUPA, Institute for Astronomy, Royal Observatory, Blackford Hill, Edinburgh, EH9 3HJ, UK \
    \and
    INAF-Istituto di Radioastronomia, Via Gobetti 101, I-40129, Bologna, Italy \
    \and
    Italian ALMA Regional Centre, Via Gobetti 101, I-40129, Bologna, Italy \
    \and
    INAF Osservatorio Astronomico di Padova, Vicolo dell'Osservatorio 5, I-35122, Padova, Italy \
    \and
    Harvard-Smithsonian Center for Astrophysics, 60 Garden St, Cambridge, MA 02138, USA \
    \and
    European Southern Observatory (ESO), Karl-Schwarzschild-Str. 2 85748 Garching bei München, Germany
    \and
    Department of Space, Earth and Environment, Chalmers University of Technology, Onsala Space Observatory, SE-439 92, Sweden \
    \and
    Leiden Observatory, Leiden University, PO Box 9513, NL-2300 RA Leiden, The Netherlands \
    \and
    Astrophysics, Department of Physics, Keble Road, Oxford, OX1 3RH, UK \
    \and
    Department of Physics \& Astronomy, University of the Western Cape, Private Bag X17,  Bellville, Cape Town, 7535, South Africa \
    \and
    Astronomy Centre, Department of Physics \& Astronomy, University of Sussex, Brighton, BN1 9QH, England \
    \and
    GEPI, Observatoire de Paris, CNRS, Université Paris Diderot, 5 place Jules Janssen, 92190 Meudon, France
    \and
    Centre for Radio Astronomy Techniques and Technologies, Department of Physics and Electronics, Rhodes University, Grahamstown 6140, South Africa
    \and
    USN, Observatoire de Paris, CNRS, PSL, UO, Nan\c cay, France
    \and
    SRON Netherlands Institute for Space Research, Landleven 12, 9747 AD, Groningen, The Netherlands \
    \and
    Kapteyn Astronomical Institute, University of Groningen, Postbus 800, 9700 AV Groningen,the Netherlands \
    }

   \date{}

 
  \abstract
   {}
   {We investigate the radio properties of a sample of 850 $\mu$m-selected sources from the SCUBA-2 Cosmology Legacy Survey (S2CLS) using new deep, low-frequency radio imaging of the Lockman Hole field from the Low Frequency Array. This sample consists of 53 sources, 41 of which are detected at $> 5 \sigma $ at 150 MHz.
   }
   {Combining these data with additional observations at 324 MHz, 610 MHz, and 1.4 GHz from the Giant Metrewave Radio Telescope and the Jansky Very Large Array, we find a variety of radio spectral shapes and luminosities ($L_{1.4 \rm{GHz}}$ ranging from $\sim4 \times 10^{23} - 1 \times 10^{25}$) within our sample despite their similarly bright submillimetre flux densities ($>4$ mJy).
   We characterise their spectral shapes in terms of multi-band radio spectral indices. Finding strong spectral flattening at low frequencies in $\sim$20\% of sources, we investigate the differences between sources with extremely flat low-frequency spectra and those with `normal' radio spectral indices ($\alpha > -0.25$). 
   }
   {As there are no other statistically significant differences between the two subgroups of our sample as split by the radio spectral index, we suggest that any differences are undetectable in galaxy-averaged properties that we can observe with our unresolved images, and likely relate to galaxy properties that we cannot resolve, on scales $\lesssim 1$ kpc.
   We attribute the observed spectral flattening in the radio to free-free absorption, proposing that those sources with significant low-frequency spectral flattening have a clumpy distribution of star-forming gas.
   We estimate an average spatial extent of absorbing material of at most several hundred parsecs to produce the levels of absorption observed in the radio spectra.
   This estimate is consistent with the highest-resolution observations of submillimetre galaxies in the literature, which find examples of non-uniform dust distributions on scales of ${\sim} 100$ pc, with evidence for clumps and knots in the interstellar medium.
Additionally, we find two bright (> 6 mJy) S2CLS sources undetected at all other wavelengths.
   We speculate that these objects may be very high redshift sources, likely residing at $z > 4$.
   }
   {}


\maketitle
%

\section{Introduction}

Bright submillimetre galaxies \citep[SMGs;][]{1997ApJ...490L...5S, 1998Natur.394..241H} are prime laboratories for investigating the physical processes involved in high redshift star formation.
With total luminosities that imply star-formation rates (SFRs) of hundreds to thousands of solar masses per year \citep[e.g.][]{1998Natur.394..248B, 2017MNRAS.469..492M} -- relative to the Milky Way's   $\sim1\,M_{\odot}~\text{yr}^{-1}$ \citep[e.g.][]{2011AJ....142..197C} -- these sources are the locations of the most extreme star formation in the Universe.
Understanding the nature of SMGs and the star formation mechanisms at play within them is an important step in beginning to address several open questions in the formation and evolution of massive galaxies.

In the local Universe, galaxies with very high infrared (IR) luminosities ($L_{\rm IR} > 10^{11}\, L_{\odot}$ and $L_{\rm IR} > 10^{12}\, L_{\odot}$ -- luminous and ultraluminous infrared galaxies, LIRGs and ULIRGs, respectively) and therefore high inferred SFRs ($> 10 - 100 \, M_{\odot} \rm{yr}^{-1}$) are rare and usually the result of recent major mergers \citep{1996ARA&A..34..749S, 1999AJ....118.2625R, 2013MNRAS.430.1901H}. 
This fits into an evolutionary picture in which mergers trigger episodes of extreme star formation, which is subsequently quenched by, for example, feedback from an active galactic nucleus (AGN), leaving a `red and dead' massive elliptical galaxy \cite[e.g.][]{2006ApJS..163....1H}.
This picture can be expanded to describe more comprehensively the diversity of star-forming galaxy populations and physical processes observed in galaxy evolution across a range of redshifts, with secular evolutionary processes also playing a role as star-formation triggers, and including galaxies with more extended, clumpy star-forming regions.
While evidence from gas kinematics suggests that at high redshift ($z > 1$) some episodes of high SFRs in galaxies may also be triggered by mergers \citep[e.g.][]{2008ApJ...680..246T}, the average SFR of the whole galaxy population is significantly higher than at low redshift \citep[e.g.][]{2014ARA&A..52..415M}.
Not only are galaxies with SFRs comparable to local LIRGs and ULIRGs the dominant galaxy population \citep{2001ApJ...556..562C, 2005ApJ...632..169L, 2013A&A...553A.132M, 2013MNRAS.429.3047B}, but a population of rare galaxies with extremely high SFRs appears at the same epoch.
The fact that such intensely star-forming galaxies are common at high redshift ($z > 1$) suggests they may play a crucial role in galaxy evolution by rapidly assembling the high stellar masses required for the formation of the most massive galaxies we see in the nearby Universe. In this evolutionary picture, SMGs at high redshift  ($z > 2$) are attractive candidates for the progenitors of massive elliptical galaxies at $z=0$ (\citealp{1999ApJ...518..641L, 2008ApJ...680..246T, 2012MNRAS.421..284H, 2014ApJ...782...68T}, but see \citealp{2020arXiv201001133G}).
Studying how this extreme star-forming phase shapes galaxy populations at high redshift is therefore crucial in uncovering the origins and nature of massive galaxy populations in the local Universe.

To gain an understanding of the mechanisms driving star formation in these galaxies, we require observations spanning a range of wavelengths, from ultraviolet (UV) to radio, to trace physical processes occurring at different energy scales.
Star-forming galaxies can emit prodigious amounts of energy from the far-infrared (FIR) to millimetre range due to the thermal heating of dust by UV photons from massive young stars \citep{1998ARA&A..36..189K}.
This re-processed emission is a robust tracer of star-formation activity in star-forming galaxies, and observations of the shape of the FIR spectral energy distribution (SED) have long been used to characterise SFRs in galaxies \citep[e.g.][]{1992ApJ...396L..69S, 1997AJ....113..599D, 2010MNRAS.409L..13C, 2012A&A...539A.155M}.

While much of the UV and optical emission is obscured in the most dusty sources, star formation may also be traced using radio continuum emission, which is not obscured by dust.
The radio spectrum arising from star formation consists of two components: thermal, free-free emission from H~{\sc ii} regions and non-thermal synchrotron emission. 
The contribution of free-free emission is only significant at high frequencies ($>10 $ GHz), and so in this study we focus on radio synchrotron emission, which dominates the spectrum at the low frequencies considered here.
This synchrotron continuum results from supernovae exploding after a delay of several megayears following the births of populations of O and B stars, which produce cosmic rays that interact with the galaxy's magnetic field.
Since both thermal FIR emission and non-thermal radio continuum emission trace physical processes associated with star formation, one would expect these quantities to correlate.
Indeed there is a well-known tight correlation between the FIR and radio luminosities of star-forming sources, the far-infrared to radio correlation \citep[FIRC;][]{1971A&A....15..110V, 1985A&A...147L...6D, 1985ApJ...298L...7H, 2010MNRAS.402..245I}, which is observed consistently across over four orders of magnitude of galaxy luminosities \citep{2001ApJ...554..803Y}.

The synchrotron radio emission of star-forming galaxies is observed to follow approximate power law behaviour at gigahertz frequencies, with a typical spectral index of $\alpha \simeq -0.7$ \citep{2013MNRAS.435..650M}\footnote{We note that the sign convention varies: In this work we use notations such that $S_{\nu} \propto \nu^{\alpha}$ and $\alpha$ is usually negative.}.
The sources, acceleration, and propagation mechanisms of the cosmic ray electrons responsible for the radio continuum emission of star-forming galaxies are still not well understood, but there are several large-scale effects that can be observed to impact the shape of the radio spectrum due to conditions of the interstellar medium (ISM).
Over time, a synchrotron spectrum will steepen at high frequencies due to radiative losses as high-energy electrons lose energy more rapidly than low-energy electrons \citep{1962SvA.....6..317K, 1968ARA&A...6..321S}.
In the low-frequency regime, free-free absorption can flatten the spectrum below a turnover frequency at which the ISM becomes optically thick \citep{1992ARA&A..30..575C, 2010MNRAS.405..887C, 2013MNRAS.431.3003L}.
Therefore, the shape of the radio synchrotron spectrum and its divergence from a simple power law hold clues as to the conditions of the ISM in and around star-forming regions.

Single-dish submillimetre observations are currently limited by the resolving power of relatively small telescopes operating at such long wavelengths, resulting in much lower-resolution observations than, for example, optical imaging.
However, observations at these wavelengths also benefit from a negative {\it K}-correction \citep{1993MNRAS.264..509B}: As galaxies are redshifted, the observed-frame 850 $\mu$m\xspace emission traces an increasingly bright part of the galaxy's FIR to millimetre spectrum due to the shape of the Rayleigh-Jeans tail, in effect compensating for cosmological dimming.
This results in the observed submillimetre flux density remaining roughly constant for a galaxy of a given luminosity between $0.5 < z < 10$ at 850 $\mu$m, enabling the detection of submillimetre sources out to very high redshifts.
Both optical and radio wavelengths, however, suffer a positive {\it K}-correction across the majority of the spectrum, meaning that the flux density of sources of the same luminosity decreases with distance.
Many studies have identified SMGs without counterparts at optical to near-infrared wavelengths \citep{2014ApJ...788..125S, 2019Natur.572..211W, 2020MNRAS.494.3828D}.
Radio observations provide a view of star formation that is not biased by dust; however, due to the positive {\it K}-correction, observing galaxies in the radio at the high redshifts at which SMGs are most numerous ($z > 2$) is very challenging.

To obtain a more complete picture of the physical processes that shape star formation in the early Universe, very deep radio surveys over wide areas of sky are required to complement deep submillimetre surveys.
Previous work has used high-resolution radio observations, typically at 1.4 GHz, as a method of pinpointing the position of submillimetre sources detected in single-dish surveys \citep{2002MNRAS.337....1I, 2005ApJ...622..772C}; however, such work has been limited by the depth of available radio sky surveys, with dedicated deep surveys over only small regions of sky, resulting in a view biased towards the brighter radio sources.
Studies have largely included limited radio spectral coverage of SMGs, focusing on the nature of the FIRC and its relation to properties such as stellar mass and redshift \citep[e.g.][]{2001ApJ...554..803Y, 2010MNRAS.402..245I, 2014MNRAS.445.2232S}.
The Low Frequency Array \citep[LOFAR;][]{2013A&A...556A...2V} has opened up new ways of studying galaxies in the radio, and a number of studies have used LOFAR's capabilities to investigate this relationship between star formation and radio luminosity in the low-frequency regime -- for example \citet{2018MNRAS.475.3010G}, \citet{2018MNRAS.480.5625R}, \citet{2020arXiv201108196S}, and \citet{2019A&A...631A.109W}.
However, these studies generally investigate the statistical properties of large samples of galaxies, in optically selected samples at low redshift (z  $\lesssim$ 2), rather than probing the shapes of individual radio spectra.
\citet{2019ApJ...883..204T} conducted an in-depth study of high-frequency (> 610 GHz) spectral curvature in SMGs, finding evidence of curved spectra that they attributed to spectral ageing of the synchrotron emission from star formation; their results implied estimated starburst ages consistent with expected SMG lifetimes.
Studies at low frequencies, where we may be able to observe absorption processes affecting the shape of the spectrum, have been hampered by a lack of sufficiently deep, wide-area data.
More comprehensive observations of the shape of the radio spectrum, extending to lower frequencies, can provide a probe of the physical conditions that give rise to extreme star formation in SMGs.
\citet{2017MNRAS.469.3468C} exploit LOFAR's frequency range to investigate the spectral shapes of star-forming galaxies and AGN, finding evidence of low-frequency spectral flattening in the star-forming sample. This sample is also constrained in redshift, focusing on local galaxies rather than the peak of star formation at z > 2, and so does not probe the bulk of the highly star-forming SMG population.
\citet{2018A&A...619A..36C} also find weak spectral flattening in local star-forming galaxies with LOFAR but largely attribute this slight effect to synchrotron losses, predicting stronger low-frequency spectral flattening due to free-free absorption at high redshift, where galaxies with high SFRs are more common.

In this study, we make use of new LOFAR deep field observations.
Reaching ${\sim}$22\,$\mu$Jy\,beam$^{-1}$, these observations have the potential to reveal the faint radio counterparts to high-redshift submillimetre sources at low frequencies.
We select a sample of SMGs using observations from the SCUBA-2 Cosmology Legacy Survey \citep[S2CLS;][]{2017MNRAS.465.1789G}, currently the largest area sky survey of its kind, which allows us to limit our study to the sites of the most extreme star formation.
Selecting sources from the Lockman Hole field, for which we have survey coverage with both LOFAR and S2CLS, we characterise their radio spectra with additional radio data from the Jansky Very Large Array (JVLA) and the Giant Metrewave Radio Telescope (GMRT).
We describe the data in Sect.~\ref{sec:data}, our sample selection in Sect.~\ref{subsec:sample}, and how we measure radio fluxes in Sect.~\ref{subsec:radio_phot}.
We briefly comment on two bright S2CLS sources that are undetected at every other wavelength in Sect.~\ref{subsec:nondetections}.
In Sects.~\ref{subsec:SEDs} onwards, we focus on the radio spectra in more detail.
We investigate the diversity of radio spectral shapes and luminosities exhibited by sources in Sect.~\ref{sec:radspec}. 

Throughout this paper we assume a flat Lambda cold dark matter ($\Lambda$CDM) cosmology with $H_0 = 69.3$\,km\,s$^{-1}$ Mpc$^{-1}$ and $\Omega_{\rm m} = 0.287$ \citep{2013ApJS..208...19H}.

\section{Data}
\label{sec:data}
    \subsection{S2CLS}
        The S2CLS observed approximately 5\,square degrees of extragalactic sky across several well-studied fields at 850 $\mu$m to a depth of ${\sim}$1\,mJy\,beam$^{-1}$, close to the SCUBA-2 confusion limit.
        In this work we focus on the Lockman Hole North field, centred at  $(\alpha,\delta)$ = 10$^{\rm h}$46$^{\rm m}$07$^{\rm s}$, $+$59$^\circ$01$'$17$''$.
        The mapping strategy resulted in an approximately circular map of diameter 30$'$, with nearly uniform noise coverage over 0.28 square degrees, at an rms depth of 1.1\,mJy\,beam$^{-1}$.
        This results in 126 submillimetre sources detected at a significance of $>4$.
        Full details of the SCUBA-2 data reduction, catalogue, and source statistics are given by \citet{2017MNRAS.465.1789G}.
        In this work we use the S2CLS source catalogue, and throughout we employ the de-boosted 850 $\mu$m flux densities. 
    
    \subsection{LOFAR}
    \label{subsec:LOFAR}

        We used the deep Lockman Hole image described by Tasse et al.\ (Paper I of the accompanying series), which is based on 112 hours of LOFAR observations.
        The image has a central rms noise level of 22 $\mu$Jy beam$^{-1}$ at a central frequency of 144 MHz and a resolution of 6 arcsec.
        Together with the images in the Bo\"otes and European Large Area Infrared Space Observatory Survey-North 1 (ELAIS-N1) fields (not covered by S2CLS) this comprises the first data release of the LOFAR Two-metre Sky Survey (LoTSS) Deep Fields and is one of the deepest images ever made at this frequency.
        It offers us the best opportunity yet available to study the low-frequency properties of distant submillimetre sources.

        Of all sources in the LOFAR catalogue, ${\sim} 98$ per cent have a candidate optical identification, selected by a combination of likelihood ratio and visual inspection using new matched-aperture multi-wavelength catalogues, as described by Kondapally et al.\ (paper III).
        For the Lockman Hole, optical data are provided by the The Spitzer Adaptation of the Red-Sequence Cluster Survey (SpARCS) and The Red Cluster Sequence Lensing Survey (RCSLenS) with the Canada France Hawaii Telescope \citep[CFHT;][]{2009ApJ...698.1943W, 2009ApJ...698.1934M, 2016MNRAS.463..635H}, and there are near-infrared data from the UKIRT Infrared Deep Sky Survey - Deep Extragalactic Survey \citep[UKIDSS DXS;][]{2007MNRAS.379.1599L} as well as mid-infrared (MIR) data from the {\it Spitzer} Wide-Area Infrared Extragalactic Survey (SWIRE) and the {\it Spitzer} Extragalactic Representative Volume Survey \citep[SERVS;][]{2003PASP..115..897L,2012PASP..124..714M}.
        Additional FIR data come from the {\it Spitzer} Multiband Imaging Photometer (MIPS) and the {\it Herschel} Multi-tiered Extragalactic Survey  \citep[HerMES;][]{2012MNRAS.424.1614O}.
        The {\it Herschel} catalogues use the optical, IR, or radio positions as a prior to obtain de-convolved flux densities for blended sources. (McCheyne et al. in prep.). 
        
        Spectroscopic redshifts were used where available, but the majority of redshifts were generated photometrically from the optical through MIR data in the manner described by Duncan et al (Paper IV).
        The broad spectral coverage in this range allows reasonable estimates of photometric redshift to be made (see Table~\ref{table:sources}).
    
    \subsection{Additional radio data}
        \label{subsec:additiona_data}
        We supplemented the LOFAR catalogue with deep archival JVLA observations at 324 MHz and 1.4 GHz \citep{2009AJ....137.4846O, 2002MNRAS.337....1I}, and at 610 MHz from the GMRT \citep{2009MNRAS.397..281I}.
        The Very Large Array (VLA) observations reach central rms noise levels of ${\sim}70\mu$Jy beam$^{-1}$ and ${\sim}4.8\mu$Jy beam$^{-1}$ at 324 MHz and 1.4 GHz, respectively, with resolutions of 6 arcsec and 1.4 arcsec.
        Observations from \citet{2009MNRAS.397..281I} at 610 MHz reach a central rms noise level of  $\sim14\mu$Jy beam$^{-1}$, with a resolution of ${\sim}6$ arcsec.
        These survey footprints cover the S2CLS Lockman Hole coverage in its entirety.
        Combined, these observations are the deepest available of the Lockman Hole field across the radio spectrum, and among the deepest radio observations to date for any extragalactic survey field.

\section{Sample selection and FIR properties}

    \subsection{Sample}
    \label{subsec:sample}
\begin{figure}
         \centering
         \includegraphics[width=\columnwidth]{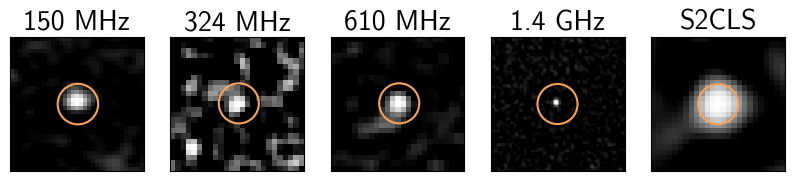}
         \caption{Cutouts of an example S2CLS source (ID 1 in our numbering system, an 11.91 mJy 850 $\mu$m source) in each radio frequency used in this study. Each square is 50 arcsec across, with the approximate S2CLS beam size (${\sim}$15 arcsec FWHM) marked with an orange circle. Image contrast is scaled arbitrarily for clarity.} 
         \label{fig:cutouts}
        \end{figure}

        Our sample consists of the 53 point sources detected at $>$5$\sigma$ -- at which significance the false detection rate falls below 1 per cent -- at 850 $\mu$m\xspace in the S2CLS Lockman Hole North field, with a median flux density of $S_{850} = 6.45$ mJy  (details in Tables~\ref{table:sources} and~\ref{table:fluxes}).
        For each submillimetre source we extracted a thumbnail image cutout in the SCUBA-2 map and at the same position in each radio map.
        Figure~\ref{fig:cutouts} shows an example source with cutouts at all four radio frequencies.
        Due to the large beam size of SCUBA-2 (${\sim}15$ arcsec full width at half maximum, FWHM), there is a risk of blending, where submillimetre flux from several galaxies that is unresolved within the large beam contributes to the source flux measurement \citep[e.g.][]{2018MNRAS.476.2278H}. 
        We checked that each S2CLS source corresponds to a single point source in the high-resolution radio images via a visual inspection of the cutout images.
        By doing so, we were able to constrain any possible multiplicity to within the 1.4 arcsec resolution of the 1.4 GHz images.
        There are a number of sources that split up into multiple components in the 1.4 GHz images, and we discuss our treatment of them below.
        
        Given the 6 arcsec resolution of the LOFAR image, we ran a simple positional cross-match between each submillimetre source and the LOFAR catalogue, identifying the closest LOFAR source within a generous 15 arcsec radius, equivalent to the size of the SCUBA-2 beam, and approximately corresponding to the 5$\sigma$ positional error on SCUBA-2 sources at the limit of the S2CLS catalogue \citep{2017MNRAS.465.1789G}. This results in 44 matched LOFAR sources and nine sources for which there is no clear LOFAR counterpart.
        We calculated the corrected Poissonian probability, $p$, of serendipitous LOFAR matches within the search radius following \citet{2011MNRAS.413.2314B} and using the LOFAR source surface density calculated above the detection threshold of the LOFAR catalogue sources.
        For identifications with $p < 0.05$, we assumed that this is a robust cross-match.
        Three identifications were found to have $p > 0.05$ and were excluded, reducing the final sample size of robust S2CLS--LOFAR matches to 41 sources.
        Only three of these sources (7 per cent) have spectroscopic redshifts available in the LOFAR catalogue,  and we used photometric redshifts where spectroscopic were not available.

        Ten of the 53 objects in the original sample break up into multiple sources in the LOFAR image such that it is unclear whether the closest positional match corresponds to the correct counterpart.
        This gives a multiplicity fraction of ${\sim} 20$ per cent.
        Estimates of multiplicity based on high-resolution Atacama Large Millimeter/submillimeter Array (ALMA) follow-up of previous single-dish submillimetre surveys range from $15 - 40$ per cent \citep[e.g.][]{2012ApJ...761...89B, 2013ApJ...776..131C, 2013ApJ...768...91H, 2017MNRAS.469..492M}, so our observed multiplicity is consistent with previous observations.
        It is possible that we missed multiples in the case that their angular separation is smaller than the resolution of our radio imaging; however, this is likely to affect only a small number of sources in our sample. 
        Of those that are visibly multiples within the S2CLS beam based on the radio images, we identified cases in which the secondary source is likely a foreground contaminant by visually inspecting cutouts of the source at radio, IR, and optical wavelengths. 
        We subsequently treated these as we did the single sources (see Sect.~\ref{subsec:radio_phot}), ensuring that the identified counterpart is identified in source extraction.
        This does neglect the case in which several sources contribute submillimetre flux and only one of which has a detectable radio counterpart, a possibility that could arise if, for example, a high redshift submillimetre source is by chance aligned with a low redshift source.
        However, our final selected sample has a median redshift of z = 2.61 (see Fig.~\ref{fig:hists} for the redshift distribution of the whole sample), which is consistent with SMG populations \citep[][Sect. 4 and references therein]{2005ApJ...622..772C, 2014PhR...541...45C}, and so it is unlikely that low redshift interlopers contaminate our sample significantly.
        There is also the possibility that the submillimetre fluxes are boosted by low luminosity sources that do not have significant radio emission -- in this case, for any significant contribution we would expect a detectable radio source given the FIRC, so this 
        also seems unlikely.
        Without high-resolution submillimetre imaging we are unable to distinguish between a single source and the above cases, and so we continue with the assumption that in all bar one of our sources we may attribute the observed submillimetre flux to a single radio source in the LOFAR catalogue.

        There are two notable sources -- sources 8 and 16 -- that are bright S2CLS detections but do not have significantly detected counterparts in any of the radio, optical, or IR images. 
        We speculate that these may be very high redshift ($z>4$) galaxies and discuss them in more detail in Sect.~\ref{subsec:nondetections}.
    \subsection{Radio fluxes}
    \label{subsec:radio_phot}
        To measure the source flux density at the four radio frequencies, we used the Source Extraction \& Photometry \citep[SEP;][]{2016JOSS....1...58B} Python package, an application of SExtractor \citep{1996A&AS..117..393B} for Python.
        We made the assumption that all sources are point-like across all radio frequencies, given that inspection of the sources at the highest angular resolution (1.4 arcesc at 1.4\,GHz) reveals no indication of resolved structure across the sample. 

        Using SEP, we measured flux densities in each of the four radio frequencies for the 41 LOFAR-detected sources described in the previous section.
        These flux densities are consistent with those presented in the LOFAR catalogue.
        We used the LOFAR catalogue coordinates from the positional cross-match with the S2CLS source list as described above as the source coordinates, given the higher angular resolution of LOFAR.
        At each radio frequency, we examined the source identification resulting from SEP and ensured that the brightest source within the SEP detection ellipse is the expected counterpart to the LOFAR position so that any multiples are not incorrectly identified as the counterpart.
        We then took the peak pixel value from within the SEP source ellipse to be the source flux density.
        Uncertainties on these measured flux densities were calculated using the off-source pixel-to-pixel rms.
       
        Of the ten sources that break up into multiple components at higher resolution, several contain sources that are likely to be foreground galaxies based on their bright optical luminosities.
        Comparing luminosities across the full range of multi-wavelength images, we were able to determine the high redshift counterpart in nine of these ten images.
        In these cases, the source SMG can be identified in the LOFAR image, and thus we analysed these as described above, using the LOFAR coordinates for reference.
        In the one case where there is no clear single counterpart and there may be truly blended emission from several co-located galaxies, further analysis would require a method of partitioning fluxes that we did not attempt, and thus we excluded the affected source from the sample.

        \begin{center}
            \begin{table*}
            \centering
            \small
            {\renewcommand{\arraystretch}{1.2}%
                \caption{Positions and redshifts of the full sample of S2CLS sources in this study. We assign IDs in column 1, which are used throughout.  We mark images where multiple sources fell in the SCUBA-2 beam size with an asterisk (*), and those that are detected at 850 $\mu$m\xspace but not in the LOFAR images we mark with a dagger ($^\dagger$). Photometric redshifts and uncertainties from the LOFAR catalogue are also provided.}
                \begin{tabular}{l l l l c}
                    \hline
                    ID  & LOFAR source ID       &       RA      &       Dec     &       z       \\  
\hline 
1       &       ILTJ104635.83+590749.2  &       $10^{\mathrm{h}}46^{\mathrm{m}}35.78^{\mathrm{s}}$      &       $+59^{\circ} 7{'}48.00{''}$  &       $1.89$$\pm^{0.71}_{0.68}$\\ 
2       &       ILTJ104644.98+591542.4  &       $10^{\mathrm{h}}46^{\mathrm{m}}45.01^{\mathrm{s}}$      &       $+59^{\circ}15{'}39.80{''}$     &       $1.30$$\pm^{0.92}_{1.08}$\\ 
3*      &       ILTJ104727.95+585213.9  &       $10^{\mathrm{h}}47^{\mathrm{m}}27.66^{\mathrm{s}}$      &       $+58^{\circ}52{'}14.60{''}$     &       $2.93$$\pm^{0.39}_{0.42}$\\ 
4       &       ILTJ104700.22+590108.1  &       $10^{\mathrm{h}}47^{\mathrm{m}}0.03^{\mathrm{s}}$       &       $+59^{\circ} 1{'}7.50{''}$   &       $2.73$$\pm^{0.48}_{0.57}$\\ 
5       &       ILTJ104535.03+585050.1  &       $10^{\mathrm{h}}45^{\mathrm{m}}35.23^{\mathrm{s}}$      &       $+58^{\circ}50{'}49.90{''}$     &       $3.76$$\pm^{0.89}_{0.86}$\\ 
6       &       ILTJ104555.38+591528.7  &       $10^{\mathrm{h}}45^{\mathrm{m}}55.19^{\mathrm{s}}$      &       $+59^{\circ}15{'}28.10{''}$     &       $4.48$$\pm^{2.72}_{2.52}$\\ 
7       &       ILTJ104632.77+590214.3  &       $10^{\mathrm{h}}46^{\mathrm{m}}32.85^{\mathrm{s}}$      &       $+59^{\circ} 2{'}12.00{''}$  &       $3.42$$\pm^{1.34}_{1.38}$\\ 
8 $^{\dagger}$  &               &       $10^{\mathrm{h}}45^{\mathrm{m}}54.58^{\mathrm{s}}$      &       $+58^{\circ}47{'}54.10{''}$     &       $  -  $ \\ 
9*      &       ILTJ104725.39+590337.8  &       $10^{\mathrm{h}}47^{\mathrm{m}}25.25^{\mathrm{s}}$      &       $+59^{\circ} 3{'}40.70{''}$  &       $2.32$$\pm^{0.74}_{0.80}$\\ 
10      &       ILTJ104631.52+585055.9  &       $10^{\mathrm{h}}46^{\mathrm{m}}31.68^{\mathrm{s}}$      &       $+58^{\circ}50{'}54.00{''}$     &       $  -  $\\ 
11      &       ILTJ104803.58+585421.4  &       $10^{\mathrm{h}}48^{\mathrm{m}}3.37^{\mathrm{s}}$       &       $+58^{\circ}54{'}22.90{''}$     &       $2.66$$\pm^{0.49}_{0.50}$\\ 
12      &       ILTJ104447.60+590035.5  &       $10^{\mathrm{h}}44^{\mathrm{m}}47.69^{\mathrm{s}}$      &       $+59^{\circ} 0{'}36.60{''}$  &       $1.98$$\pm^{0.69}_{0.78}$\\ 
13      &       ILTJ104720.51+591043.6  &       $10^{\mathrm{h}}47^{\mathrm{m}}20.57^{\mathrm{s}}$      &       $+59^{\circ}10{'}40.90{''}$     &       $2.69$$\pm^{0.80}_{0.66}$\\ 
14      &       ILTJ104657.32+591459.2  &       $10^{\mathrm{h}}46^{\mathrm{m}}57.26^{\mathrm{s}}$      &       $+59^{\circ}14{'}57.60{''}$     &       $2.96$$\pm^{0.75}_{0.62}$\\ 
15      &       ILTJ104456.67+585000.0  &       $10^{\mathrm{h}}44^{\mathrm{m}}56.86^{\mathrm{s}}$      &       $+58^{\circ}49{'}59.00{''}$     &       $  -  $ \\ 
16 $^{\dagger}$ &               &       $10^{\mathrm{h}}45^{\mathrm{m}}1.83^{\mathrm{s}}$       &       $+59^{\circ} 4{'}3.10{''}$   &       $  -  $ \\ 
17*     &       ILTJ104717.95+590232.2  &       $10^{\mathrm{h}}47^{\mathrm{m}}18.21^{\mathrm{s}}$      &       $+59^{\circ} 2{'}31.00{''}$  &       $2.31$$\pm^{0.34}_{0.39}$\\ 
18      &       ILTJ104556.88+585318.8  &       $10^{\mathrm{h}}45^{\mathrm{m}}56.87^{\mathrm{s}}$      &       $+58^{\circ}53{'}18.10{''}$     &       $1.39$$\pm^{0.64}_{0.68}$\\ 
19      &       ILTJ104702.46+585102.9  &       $10^{\mathrm{h}}47^{\mathrm{m}}2.61^{\mathrm{s}}$       &       $+58^{\circ}51{'}5.40{''}$      &       $3.28$$\pm^{1.89}_{1.86}$\\ 
20      &       ILTJ104800.86+590343.8  &       $10^{\mathrm{h}}48^{\mathrm{m}}1.05^{\mathrm{s}}$       &       $+59^{\circ} 3{'}43.10{''}$  &       $  -  $ \\ 
21      &       ILTJ104633.12+585158.7  &       $10^{\mathrm{h}}46^{\mathrm{m}}33.24^{\mathrm{s}}$      &       $+58^{\circ}52{'}0.00{''}$      &       $3.00$$\pm^{1.16}_{1.14}$\\ 
22*     &       ILTJ104523.51+591631.2  &       $10^{\mathrm{h}}45^{\mathrm{m}}23.87^{\mathrm{s}}$      &       $+59^{\circ}16{'}25.70{''}$     &       $0.82$$\pm^{0.20}_{0.16}$\\ 
23      &       ILTJ104351.14+590058.1  &       $10^{\mathrm{h}}43^{\mathrm{m}}51.48^{\mathrm{s}}$      &       $+59^{\circ} 0{'}57.70{''}$  &       $2.27$$\pm^{0.56}_{0.60}$\\ 
24*     &       ILTJ104626.25+590539.5  &       $10^{\mathrm{h}}46^{\mathrm{m}}26.92^{\mathrm{s}}$      &       $+59^{\circ} 5{'}44.10{''}$  &       $  -  $ \\ 
25      &       ILTJ104440.17+585929.8  &       $10^{\mathrm{h}}44^{\mathrm{m}}40.23^{\mathrm{s}}$      &       $+58^{\circ}59{'}28.30{''}$     &       $2.14$$\pm^{0.74}_{1.37}$\\ 
26*     &       ILTJ104715.52+590636.6  &       $10^{\mathrm{h}}47^{\mathrm{m}}15.49^{\mathrm{s}}$      &       $+59^{\circ} 6{'}33.10{''}$  &       $3.95$$\pm^{2.06}_{2.14}$\\ 
27*$^{\dagger}$ &               &       $10^{\mathrm{h}}47^{\mathrm{m}}20.94^{\mathrm{s}}$      &       $+58^{\circ}51{'}52.90{''}$     &       $  -  $ \\ 
28      &       ILTJ104633.11+591220.2  &       $10^{\mathrm{h}}46^{\mathrm{m}}32.97^{\mathrm{s}}$      &       $+59^{\circ}12{'}20.00{''}$     &       $2.61$$\pm^{1.19}_{1.09}$\\ 
29      &       ILTJ104522.33+591726.0  &       $10^{\mathrm{h}}45^{\mathrm{m}}22.55^{\mathrm{s}}$      &       $+59^{\circ}17{'}21.70{''}$     &       $1.99$$\pm^{0.76}_{0.74}$\\ 
30 $^{\dagger}$ &               &       $10^{\mathrm{h}}48^{\mathrm{m}}0.04^{\mathrm{s}}$       &       $+58^{\circ}54{'}47.10{''}$     &       $  -  $ \\ 
31      &       ILTJ104813.45+590340.9  &       $10^{\mathrm{h}}48^{\mathrm{m}}13.49^{\mathrm{s}}$      &       $+59^{\circ} 3{'}38.30{''}$  &       $3.19$$\pm^{2.12}_{2.14}$\\ 
32*     &       ILTJ104630.75+585908.3  &       $10^{\mathrm{h}}46^{\mathrm{m}}31.00^{\mathrm{s}}$      &       $+58^{\circ}59{'}8.00{''}$      &       $5.64$$\pm^{1.17}_{1.08}$\\ 
33      &       ILTJ104734.49+591333.2  &       $10^{\mathrm{h}}47^{\mathrm{m}}34.22^{\mathrm{s}}$      &       $+59^{\circ}13{'}28.40{''}$     &       $3.15$$\pm^{1.97}_{2.06}$\\ 
34*     &       ILTJ104718.16+585525.9  &       $10^{\mathrm{h}}47^{\mathrm{m}}18.23^{\mathrm{s}}$      &       $+58^{\circ}55{'}25.00{''}$     &       $3.84$$\pm^{1.23}_{1.34}$\\ 
35      &       ILTJ104638.62+585612.6  &       $10^{\mathrm{h}}46^{\mathrm{m}}38.72^{\mathrm{s}}$      &       $+58^{\circ}56{'}11.90{''}$     &       $2.27$$\pm^{0.32}_{0.33}$\\ 
36      &       ILTJ104700.07+585441.5  &       $10^{\mathrm{h}}46^{\mathrm{m}}59.87^{\mathrm{s}}$      &       $+58^{\circ}54{'}37.50{''}$     &       $  -  $ \\ 
37 $^{\dagger}$ &               &       $10^{\mathrm{h}}46^{\mathrm{m}}23.24^{\mathrm{s}}$      &       $+58^{\circ}59{'}36.10{''}$     &       $  -  $ \\ 
38      &       ILTJ104822.99+590112.1  &       $10^{\mathrm{h}}48^{\mathrm{m}}22.92^{\mathrm{s}}$      &       $+59^{\circ} 1{'}9.70{''}$   &       $  -  $ \\ 
39      &       ILTJ104431.34+590612.8  &       $10^{\mathrm{h}}44^{\mathrm{m}}31.37^{\mathrm{s}}$      &       $+59^{\circ} 6{'}15.90{''}$  &       $1.14$$\pm^{0.16}_{0.15}$\\ 
40 $^{\dagger}$ &               &       $10^{\mathrm{h}}44^{\mathrm{m}}42.19^{\mathrm{s}}$      &       $+59^{\circ} 2{'}10.40{''}$  &       $  -  $ \\ 
41      &       ILTJ104608.72+585828.7  &       $10^{\mathrm{h}}46^{\mathrm{m}}8.49^{\mathrm{s}}$       &       $+58^{\circ}58{'}28.20{''}$     &       $2.09$$\pm^{0.90}_{0.97}$\\ 
42      &       ILTJ104430.59+585518.4  &       $10^{\mathrm{h}}44^{\mathrm{m}}30.59^{\mathrm{s}}$      &       $+58^{\circ}55{'}15.90{''}$     &       $1.94$$\pm^{0.41}_{0.36}$\\ 
43      &       ILTJ104730.66+590427.5  &       $10^{\mathrm{h}}47^{\mathrm{m}}30.73^{\mathrm{s}}$      &       $+59^{\circ} 4{'}22.50{''}$  &       $1.76$$\pm^{0.69}_{0.68}$\\ 
44      &       ILTJ104601.72+590917.3  &       $10^{\mathrm{h}}46^{\mathrm{m}}1.99^{\mathrm{s}}$       &       $+59^{\circ} 9{'}18.20{''}$  &       $1.88$$\pm^{0.29}_{0.28}$\\ 
45      &       ILTJ104731.18+591134.4  &       $10^{\mathrm{h}}47^{\mathrm{m}}29.72^{\mathrm{s}}$      &       $+59^{\circ}11{'}32.50{''}$     &       $  -  $ \\ 
46      &       ILTJ104601.56+585153.4  &       $10^{\mathrm{h}}46^{\mathrm{m}}1.26^{\mathrm{s}}$       &       $+58^{\circ}51{'}52.20{''}$     &       $2.27$$\pm^{0.92}_{1.36}$\\ 
47 $^{\dagger}$ &               &       $10^{\mathrm{h}}46^{\mathrm{m}}21.39^{\mathrm{s}}$      &       $+58^{\circ}54{'}34.10{''}$     &       $  -  $ \\ 
48 $^{\dagger}$ &               &       $10^{\mathrm{h}}44^{\mathrm{m}}18.32^{\mathrm{s}}$      &       $+59^{\circ} 2{'}41.30{''}$  &       $  -  $ \\ 
49      &       ILTJ104744.66+591413.6  &       $10^{\mathrm{h}}47^{\mathrm{m}}44.16^{\mathrm{s}}$      &       $+59^{\circ}14{'}11.90{''}$     &       $2.20$$\pm^{0.58}_{0.53}$\\ 
50      &       ILTJ104444.87+591500.9  &       $10^{\mathrm{h}}44^{\mathrm{m}}44.53^{\mathrm{s}}$      &       $+59^{\circ}14{'}50.50{''}$     &       $  -  $ \\ 
51*     &       ILTJ104354.98+590616.7  &       $10^{\mathrm{h}}43^{\mathrm{m}}55.28^{\mathrm{s}}$      &       $+59^{\circ} 6{'}16.00{''}$  &       $2.84$$\pm^{0.86}_{0.82}$\\ 
52      &       ILTJ104539.62+584829.8  &       $10^{\mathrm{h}}45^{\mathrm{m}}39.90^{\mathrm{s}}$      &       $+58^{\circ}48{'}30.00{''}$     &       $3.70$$\pm^{1.45}_{1.45}$\\ 
53      &       ILTJ104738.02+585634.2  &       $10^{\mathrm{h}}47^{\mathrm{m}}37.65^{\mathrm{s}}$      &       $+58^{\circ}56{'}32.20{''}$     &       $  -  $ \\ 
\hline 

                \end{tabular}
                \label{table:sources}
                }
            
            \end{table*}
        \end{center}
    
         \begin{center}
            \begin{table*}
            \centering
            \small
            {\renewcommand{\arraystretch}{1.2}%
                \caption{Fluxes and radio spectral indices of sources in this study. IDs in column 1 follow Table~\ref{table:sources}, with multiple sources marked with an asterisk (*) and those not detected in LOFAR with a dagger ($^\dagger$). De-boosted 850 $\mu$m\xspace flux densities are shown from the S2CLS catalogue, and radio flux densities are measured as described in Sect.~\ref{subsec:radio_phot}. Here, $\alpha_{150-324}$ and  $\alpha_{324-1400}$ are the low- and high-frequency radio spectral indices, respectively, as described in Sect.~\ref{sec:radspec}.}
                \begin{tabular}{l r r r r r r r}
                    \hline
                   ID   &       S$_{850} $      &       S$_{150 \rm{MHz}} $       &       S$_{324 \rm{MHz}} $     &       S$_{610 \rm{MHz}} $     &       S$_{1.4 \rm{GHz}} $     &       $\alpha_{150-324}$      &       $\alpha_{324-1400}$                     \\  
                &       (mJy)           &       ($\mu$Jy)                       &       ($\mu$Jy)                       &       ($\mu$Jy)                       &       ($\mu$Jy)                       &                                               &                                                               \\  
\hline 
1       &       11.91$\pm$ 1.23 &       $515.87\pm 21.88$       &       $269.05\pm 74.19$  &       $215.36\pm 16.93$       &       $80.64\pm 3.72$ &       $-0.81$ &       $-0.82$\\ 
2       &       12.28$\pm$ 0.00 &       $569.90\pm 45.00$       &       $326.56\pm 69.59$  &       $230.62\pm 18.62$       &       $60.12\pm 6.04$ &       $-0.69$ &       $-1.16$\\ 
3*      &       9.91$\pm$ 1.33  &       $437.02\pm 84.05$       &       $218.79\pm 69.63$  &       $243.83\pm 34.52$       &       $76.62\pm 6.97$ &       $-0.86$ &       $-0.72$\\ 
4       &       8.92$\pm$ 1.62  &       $627.80\pm 26.03$       &       $654.40\pm 67.81$  &       $463.67\pm 12.53$       &       $232.95\pm 3.45$        &       $0.05$  &       $-0.71$\\ 
5       &       8.66$\pm$ 1.43  &       $260.40\pm 33.58$       &       $185.44\pm 65.30$  &       $97.98\pm 16.65$        &       $46.60\pm 4.08$ &       $-0.42$ &       $-0.94$\\ 
6       &       9.68$\pm$ 0.91  &       $155.09\pm 28.09$       &       $226.77\pm 63.90$  &       $66.25\pm 14.25$        &       $32.53\pm 4.82$ &       $0.47$  &       $-1.33$\\ 
7       &       8.15$\pm$ 1.28  &       $427.66\pm 30.85$       &       $318.65\pm 81.14$  &       $217.56\pm 15.91$       &       $119.51\pm 3.12$        &       $-0.37$ &       $-0.67$\\ 
8 $^{\dagger}$  &       8.31$\pm$ 1.50  &       $53.22\pm 217.11$       &       $-73.58\pm 306.52$ &       $39.84\pm 37.60$        &       $-21.12\pm 65.69$       &       $  -  $   &       $  -  $\\ 
9*      &       7.92$\pm$ 1.37  &       $352.32\pm 24.68$       &       $242.35\pm 45.25$  &       $110.80\pm 11.23$       &       $36.95\pm 3.90$ &       $-0.47$ &       $-1.29$\\ 
10      &       7.91$\pm$ 1.36  &       $2757.32\pm 29.80$      &       $1591.78\pm 65.92$  &       $923.99\pm 10.53$       &       $265.02\pm 4.33$        &       $-0.68$ &       $-1.23$\\ 
11      &       8.92$\pm$ 1.35  &       $2901.79\pm 33.26$      &       $1886.64\pm 66.49$  &       $965.98\pm 8.66$        &       $261.03\pm 6.77$        &       $-0.54$ &       $-1.35$\\ 
12      &       7.45$\pm$ 1.32  &       $891.28\pm 35.67$       &       $597.49\pm 64.44$  &       $314.52\pm 14.95$       &       $131.48\pm 4.16$        &       $-0.50$ &       $-1.03$\\ 
13      &       7.35$\pm$ 1.33  &       $677.98\pm 25.79$       &       $399.62\pm 75.91$  &       $218.91\pm 11.76$       &       $65.99\pm 4.84$ &       $-0.66$ &       $-1.23$\\ 
14      &       7.87$\pm$ 0.86  &       $382.77\pm 22.14$       &       $211.55\pm 78.58$  &       $173.30\pm 15.20$       &       $34.05\pm 6.29$ &       $-0.74$ &       $-1.25$\\ 
15      &       7.56$\pm$ 1.40  &       $195.09\pm 82.58$       &       $-142.94\pm 252.32$ &       $76.64\pm 64.94$        &       $202.57\pm 69.16$       &       $  -  $   &       $  -  $\\ 
16 $^{\dagger}$ &       6.81$\pm$ 1.44  &       $147.40\pm 187.52$      &       $-136.26\pm 286.97$ &       $39.09\pm 33.00$        &       $-115.27\pm 52.75$      &       $  -  $   &       $  -  $\\ 
17*     &       6.77$\pm$ 1.37  &       $396.25\pm 18.82$       &       $230.10\pm 75.84$  &       $155.23\pm 13.57$       &       $39.92\pm 3.66$ &       $-0.68$ &       $-1.20$\\ 
18      &       6.76$\pm$ 1.42  &       $786.16\pm 36.80$       &       $611.55\pm 80.54$  &       $413.93\pm 14.34$       &       $205.57\pm 4.10$        &       $-0.31$ &       $-0.74$\\ 
19      &       6.72$\pm$ 1.30  &       $257.25\pm 21.91$       &       $304.00\pm 75.83$  &       $68.13\pm 10.48$        &       $31.74\pm 4.65$ &       $0.21$  &       $-1.54$\\ 
20      &       7.45$\pm$ 1.14  &       $201.01\pm 117.18$      &       $63.77$ &       $67.83\pm 48.35$  &       $-243.98\pm 90.86$      &       $  -  $ &       $  -  $\\ 
21      &       6.45$\pm$ 1.36  &       $246.71\pm 25.95$       &       $170.03\pm 75.18$  &       $129.69\pm 9.11$        &       $55.18\pm 3.85$ &       $-0.46$ &       $-0.77$\\ 
22*     &       8.25$\pm$ 1.51  &       $599.27\pm 68.02$       &       $349.86\pm 88.50$  &       $187.96\pm 53.06$       &       $30.08\pm 8.89$ &       $-0.67$ &       $-1.68$\\ 
23      &       8.15$\pm$ 2.06  &       $306.75\pm 21.86$       &       $302.24\pm 70.87$  &       $123.97\pm 11.61$       &       $36.63\pm 6.55$ &       $-0.02$ &       $-1.44$\\ 
24*     &       6.47$\pm$ 1.19  &       $611.65\pm 412.64$      &       $46.51\pm 349.55$ &       $179.32\pm 186.81$      &       $286.13\pm 50.29$       &       $  -  $   &       $  -  $\\ 
25      &       6.31$\pm$ 1.36  &       $514.89\pm 26.75$       &       $248.83\pm 67.93$  &       $191.35\pm 9.75$        &       $67.50\pm 3.82$ &       $-0.90$ &       $-0.89$\\ 
26*     &       6.50$\pm$ 1.39  &       $316.26\pm 44.95$       &       $200.46\pm 74.80$  &       $148.73\pm 19.29$       &       $45.82\pm 3.71$ &       $-0.57$ &       $-1.01$\\ 
27*$^{\dagger}$ &       6.32$\pm$ 1.57  &       $331.03\pm 103.48$      &       $314.05\pm 261.99$ &       $31.25\pm 55.27$        &       $456.51\pm 75.47$       &       $  -  $   &       $  -  $\\ 
28      &       6.17$\pm$ 1.27  &       $280.43\pm 29.52$       &       $134.93\pm 53.37$  &       $122.89\pm 12.37$       &       $47.90\pm 4.63$ &       $-0.91$ &       $-0.71$\\ 
29      &       8.84$\pm$ 1.68  &       $417.91\pm 25.87$       &       $208.33\pm 55.58$  &       $196.16\pm 14.54$       &       $55.21\pm 6.88$ &       $-0.87$ &       $-0.91$\\ 
30 $^{\dagger}$ &       6.42$\pm$ 1.52  &       $503.76\pm 114.80$      &       $781.95\pm 283.17$ &       $153.31\pm 75.00$       &       $420.91\pm 93.98$       &       $  -  $   &       $  -  $\\ 
31      &       6.69$\pm$ 1.53  &       $214.44\pm 18.08$       &       $203.56\pm 82.41$  &       $77.61\pm 16.35$        &       $25.53\pm 6.76$ &       $-0.06$ &       $-1.42$\\ 
32*     &       5.52$\pm$ 1.21  &       $197.99\pm 37.48$       &       $186.59\pm 115.89$ &       $105.71\pm 69.49$       &       $32.39\pm 3.22$ &       $-0.07$ &       $-1.20$\\ 
33      &       6.30$\pm$ 1.33  &       $320.15\pm 33.35$       &       $182.35\pm 47.71$  &       $111.52\pm 17.30$       &       $45.49\pm 6.72$ &       $-0.70$ &       $-0.95$\\ 
34*     &       5.44$\pm$ 1.30  &       $224.34\pm 52.19$       &       $245.03\pm 77.06$  &       $56.77\pm 21.37$        &       $39.70\pm 5.24$ &       $0.11$  &       $-1.24$\\ 
35      &       5.50$\pm$ 1.33  &       $603.71\pm 32.53$       &       $367.44\pm 71.91$  &       $283.81\pm 15.07$       &       $129.00\pm 3.86$        &       $-0.62$ &       $-0.72$\\ 
36      &       5.42$\pm$ 1.18  &       $799.27\pm 254.92$      &       $164.88\pm 326.85$ &       $329.11\pm 256.36$      &       $368.81\pm 58.78$       &       $  -  $   &       $  -  $\\ 
37 $^{\dagger}$ &       4.86$\pm$ 1.29  &       $520.94\pm 83.10$       &       $-29.62\pm 313.97$ &       $136.86\pm 60.50$       &       $130.38\pm 48.72$       &       $  -  $   &       $  -  $\\ 
38      &       6.37$\pm$ 1.38  &       $748.04\pm 128.77$      &       $232.76$        &       $45.05\pm 53.30$  &       $124.80\pm 119.67$      &       $  -  $ &       $  -  $\\ 
39      &       4.98$\pm$ 1.33  &       $290.25\pm 33.77$       &       $315.02\pm 75.94$  &       $153.90\pm 8.45$        &       $53.63\pm 4.57$ &       $0.10$  &       $-1.21$\\ 
40 $^{\dagger}$ &       5.07$\pm$ 1.20  &       $67.59\pm 98.51$        &       $-341.11\pm 221.77$ &       $99.69\pm 42.18$        &       $20.18\pm 59.61$        &       $  -  $   &       $  -  $\\ 
41      &       5.05$\pm$ 1.31  &       $1276.44\pm 27.34$      &       $786.63\pm 65.59$  &       $447.18\pm 12.20$       &       $189.72\pm 2.92$        &       $-0.60$ &       $-0.97$\\ 
42      &       5.25$\pm$ 1.39  &       $475.66\pm 28.96$       &       $181.98\pm 63.07$  &       $192.92\pm 8.36$        &       $44.27\pm 4.53$ &       $-1.20$ &       $-0.97$\\ 
43      &       5.02$\pm$ 1.28  &       $341.40\pm 26.43$       &       $331.18\pm 66.97$  &       $162.85\pm 11.18$       &       $65.27\pm 4.43$ &       $-0.04$ &       $-1.11$\\ 
44      &       4.73$\pm$ 1.19  &       $464.62\pm 40.36$       &       $345.32\pm 85.46$  &       $193.13\pm 10.54$       &       $64.20\pm 3.66$ &       $-0.37$ &       $-1.15$\\ 
45      &       5.31$\pm$ 1.43  &       $76.03\pm 239.46$       &       $-19.76\pm 232.55$ &       $164.37\pm 84.16$       &       $139.95\pm 95.10$       &       $  -  $   &       $  -  $\\ 
46      &       4.53$\pm$ 1.14  &       $516.47\pm 25.07$       &       $297.33\pm 74.08$  &       $222.81\pm 9.71$        &       $116.54\pm 3.86$        &       $-0.69$ &       $-0.64$\\ 
47 $^{\dagger}$ &       4.66$\pm$ 1.15  &       $-106.75\pm 81.23$      &       $32.22\pm 250.29$ &       $-10.28\pm 49.18$       &       $93.86\pm 52.67$        &       $  -  $   &       $  -  $\\ 
48 $^{\dagger}$ &       5.00$\pm$ 1.51  &       $271.83\pm 80.87$       &       $-131.63$       &       $60.10\pm 51.23$  &       $-119.05\pm 67.70$      &       $  -  $ &       $  -  $\\ 
49      &       5.89$\pm$ 1.28  &       $197.55\pm 24.97$       &       $150.22\pm 68.25$  &       $104.23\pm 16.23$       &       $40.82\pm 7.98$ &       $-0.34$ &       $-0.89$\\ 
50      &       6.40$\pm$ 1.59  &       $452.53\pm 200.45$      &       $174.61\pm 244.19$ &       $173.99\pm 69.35$       &       $49.36\pm 105.88$       &       $  -  $   &       $  -  $\\ 
51*     &       6.04$\pm$ 1.76  &       $332.04\pm 66.28$       &       $231.01\pm 91.76$  &       $145.52\pm 40.04$       &       $51.86\pm 8.26$ &       $-0.45$ &       $-1.02$\\ 
52      &       4.60$\pm$ 1.30  &       $323.68\pm 27.13$       &       $252.34\pm 69.15$  &       $187.14\pm 11.54$       &       $67.46\pm 4.56$ &       $-0.31$ &       $-0.90$\\ 
53      &       4.59$\pm$ 1.19  &       $402.38\pm 97.36$       &       $43.27\pm 273.87$ &       $61.92\pm 42.72$        &       $-30.77\pm 74.99$       &       $  -  $   &       $  -  $\\ 
\hline 

                \end{tabular}
                
            \label{table:fluxes}
                }
            \end{table*}
        \end{center}

    \subsection{Non-detections}
    \label{subsec:nondetections}
    
\begin{figure}
         \centering
         \includegraphics[width=\columnwidth]{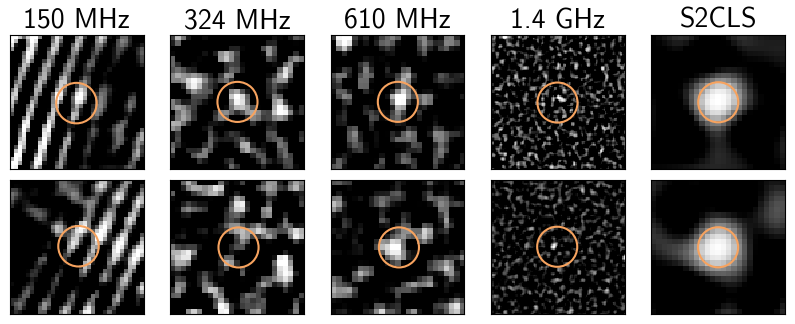}
         \caption{Cutouts of sources 8 and 16 at radio frequencies, with S2CLS 850 $\mu$m\xspace. There are clear artefacts in the LOFAR image, which result from nearby bright sources. While low S/N detections look possible in several bands, these sources lie below the 5$\sigma$ detection threshold in all bands other than 850 $\mu$m\xspace. The orange circle shows the 15 arcsec SCUBA-2 beam FWHM.}
         \label{fig:cutout8}
        \end{figure}

        Two S2CLS detections -- S2CLSJ104554+584754 and S2CLSJ104501+590403, sources 8 and 16, respectively, in our sample -- are below the $5\sigma$ detection threshold in the LOFAR map.
        Inspecting these sources in all other available wavelengths, we also find them to be below the $5\sigma$ detection thresholds across the optical and IR as well as at all other radio frequencies (see Fig.~\ref{fig:cutout8}).

        While eight sources appear in the S2CLS catalogue that do not have counterparts in the LOFAR catalogue, these two sources are of particular interest due to their high S/Ns.
        With S/Ns of 8.14 and 7.14, respectively, they are very unlikely to be false detections, the false detection rate in S2CLS being less than 0.001 per cent for a $\geqslant 7 \sigma$ source.
        We focused on these two bright S2CLS sources as they are most likely to be real sources, and we did not attempt a further analysis of the fainter sources without significant multi-wavelength counterparts due to the probability of them being false detections.

        A number of possible factors could lead to these bright submillimetre sources with very faint fluxes at other wavelengths: It is possible that these submillimetre detections are the result of blending of several faint sources with an angular separation smaller than the SCUBA-2 beam but larger than at other wavelengths.
        However, if we assume that these are indeed single sources, then these objects are either extremely reddened or they are at very high redshift ($z > 4$).
        Most likely, we are observing a combination of both redshift and reddening effects \citep{1998Natur.394..241H, 2012Natur.486..233W}, and thus these lone SCUBA-2 detections are candidate high redshift sources.

        To estimate a limiting lower redshift at which these sources must be located to be detected at 850 $\mu$m\xspace and below the detection threshold at other wavelengths given the depths of the various surveys, we determined the detection thresholds in each of the images in which there is no detection.
        For the optical through to near infrared, we used 5$\sigma$ detection thresholds from position-dependent depth maps \citep{2019MNRAS.490..634S}.
        For the {\it Herschel} Photodetector Array Camera and Spectrometer (PACS) and Spectral and Photometric Imaging Receive (SPIRE) images, we used the documented 5$\sigma$ survey depths from \citet{2012MNRAS.424.1614O}.
        For Spitzer MIPS, we used the survey depths from the SWIRE second data release \citep[DR2;][]{2012yCat.2302....0S} \footnote{Values for survey depths can be found at the following URL: \url{http://spider.ipac.caltech.edu/staff/jason/swire/astronomers/program.html}}.

\begin{figure*}
        \centering
         \includegraphics[width=\textwidth]{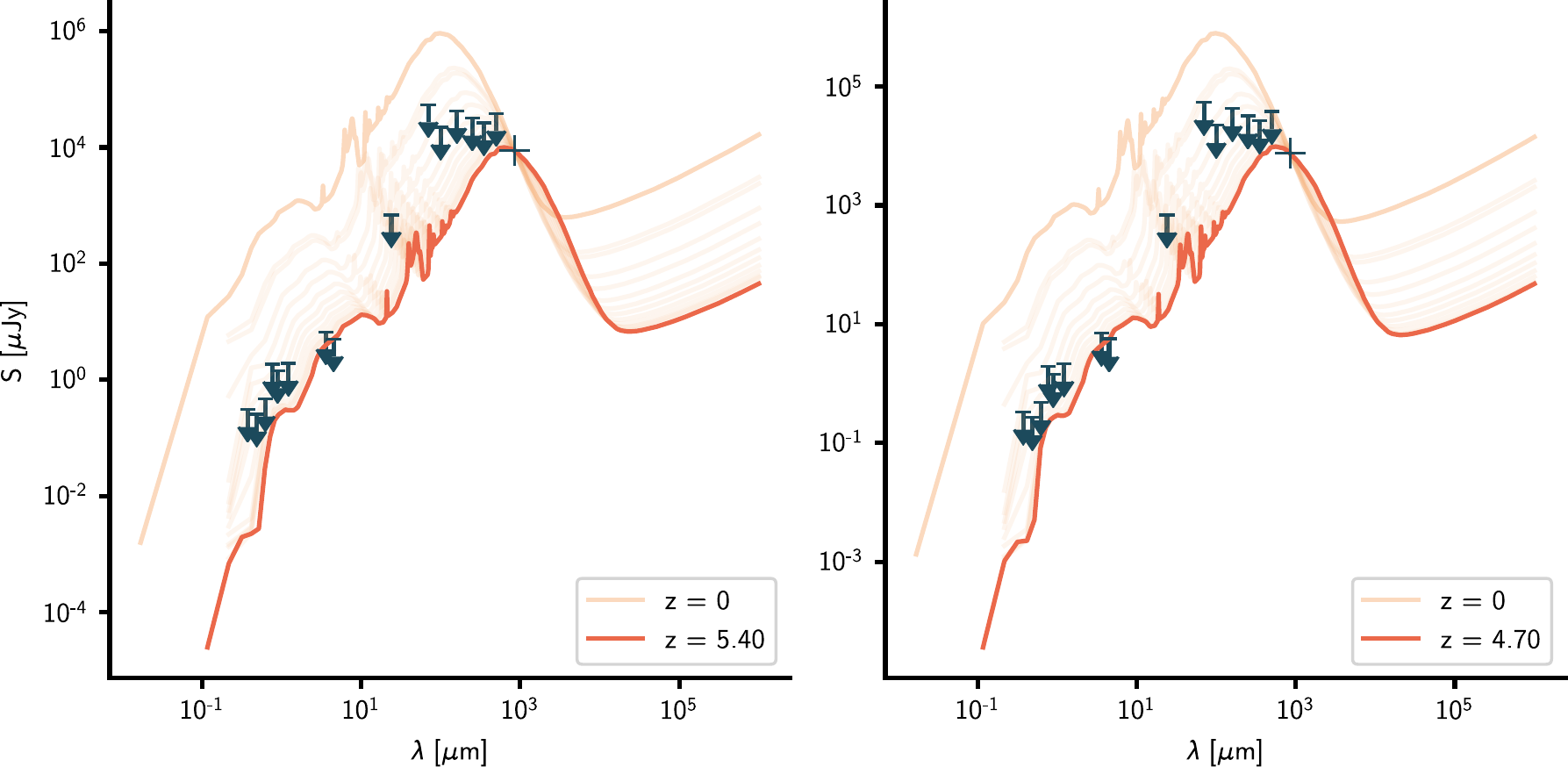}
         \caption{Calculating the lower limits on redshifts for sources 8 (left) and 16 (right), both of which are detected at 850 $\mu$m\xspace and at no other wavelength in this study. Using the $5\sigma$ detection thresholds (marked as downward pointing arrows) at optical through to FIR wavelengths, and the detection at 850 $\mu$m\xspace, we redshift the \citet{2010A&A...514A..67M} template SED until it lies below the limiting fluxes. In this way we determine limits of $z = 5.4$ and $z = 4.7$ for sources 8 and 16, respectively.}
         \label{fig:redshift8}
        \end{figure*}

        As a simple illustration, we used the averaged SMG template SED from \citet{2010A&A...514A..67M} as an assumed, underlying SED for these two sources.
        Normalising at the observed 850 $\mu$m\xspace flux density, we redshifted the template until it lay below the detection thresholds of each wavelength (see Fig.~\ref{fig:redshift8}).
        Thus we obtained a lower limit on the redshift of each source.

        This results in limiting redshifts of $z = 5.4$ for source 8 and $z = 4.7$ for source 16, both of which are plausible for highly star-forming galaxies selected at 850 $\mu$m.
        Given the simplistic nature of our fitting procedure, and the assumptions about the underlying SED shape, we cannot give robust redshift limits on these sources;  however, this is an indication of an approximate redshift that is consistent with our hypothesis that these are two submillimetre-bright, optically and radio-dim sources located at high redshift.
        Further investigation of these sources with submillimetre or millimetre instruments to obtain spectral line emission would be required to more accurately determine the redshifts of these objects.
        
    \subsection{Optical to submillimetre spectral energy distributions}
    \label{subsec:SEDs}
        As described in Sect.~\ref{subsec:radio_phot}, in the majority of cases there is a clear, single LOFAR counterpart to the submillimetre source. Therefore, to construct the SED we can simply use the LOFAR catalogue source and the corresponding flux densities and uncertainties at all optical and IR wavelengths provided in the LOFAR added-value catalogue (Kondapally et al., 2020, paper III in series) and the radio flux densities measured using the method described in Sect.~\ref{subsec:radio_phot}.

\begin{figure*}
        \centering
     \includegraphics[width=\textwidth]{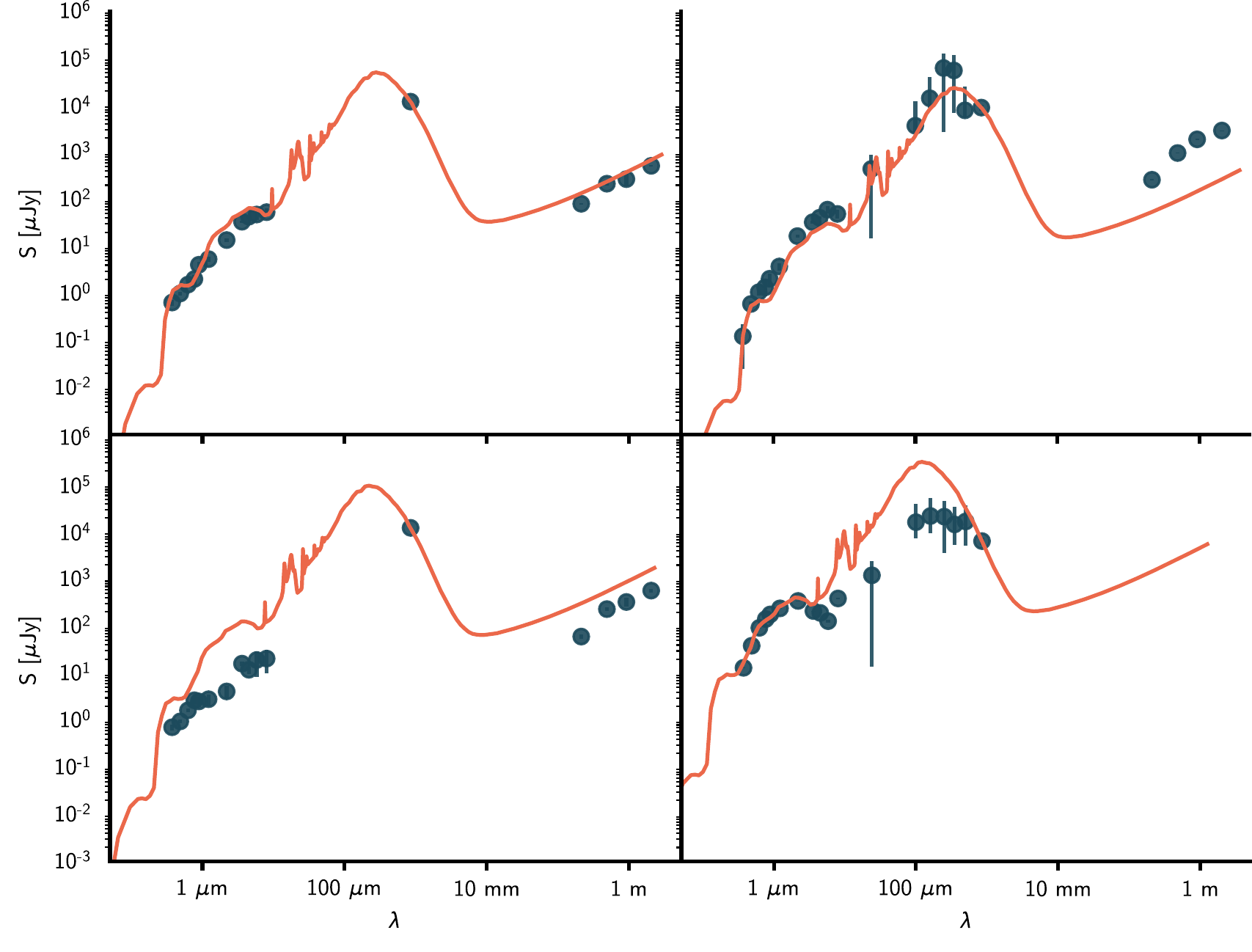}
     \caption{Observed-frame SEDs of four example sources from our sample. Blue points show the flux densities across the spectrum, with radio flux densities measured as described in Sect.~\ref{subsec:radio_phot} and optical--FIR from the LOFAR catalogue. Uncertainties are plotted, though in many cases they are smaller than the size of the plotted points. The average SMG template from \citet{2010A&A...514A..67M} (orange) is over-plotted, normalised at the observed-frame 850 $\mu$m\xspace SCUBA-2 data point. We show several examples to demonstrate that, in most cases, the template traces the IR dust spectrum well (though in others, less well) but that due to the lack of {\it Herschel} data covering the peak of the dust spectrum in several sources, we use this normalised template to estimate FIR luminosities instead of template fitting.     }
     \label{fig:SED}
    \end{figure*}

        Figure~\ref{fig:SED} shows an example SED for a source in our sample. 
        We over-plot the \citet{2010A&A...514A..67M} template SED normalised at the  observed frame 850 $\mu$m\xspace flux density and transformed to the photometric redshift of each source.
        This provides a visual comparison to a typical high redshift dusty star-forming galaxy SED, and we can see that the SEDs of sources in this study are broadly similar across the full wavelength range.
        As several sources in our sample are outside of {\it Herschel} survey area coverage, and therefore do not have data available with which to constrain the peak of dust spectrum, we did not conduct a full SED-fitting process but simply used the normalised star-forming galaxy templates.
        While this is not as robust as a fit to the observed SEDs, in the majority of cases the template traces the observed data adequately. 

    \subsection{The far-infrared to radio correlation (FIRC)}
    \label{subsec:FIRC}
        As a quick check to confirm that none of our sources stand out as radio-loud AGN, we calculated estimates of FIR luminosity for our sources and plotted them on the FIRC. 
        We used a very simple method, integrating the \citet{2010A&A...514A..67M} average SMG template flux between 8 and 1000 $\mu$m\xspace in the rest frame.
        We then normalised this against the observed-frame 850 $\mu$m\xspace luminosity.
        While this is a naive approach and inevitably introduces some uncertainty, the \citet{2010A&A...514A..67M} template SED is qualitatively similar enough across the wavelength range in consideration to provide a reasonable estimate of the FIR luminosity of the sources.
        
\begin{figure}
            \centering
            \includegraphics[width=\columnwidth]{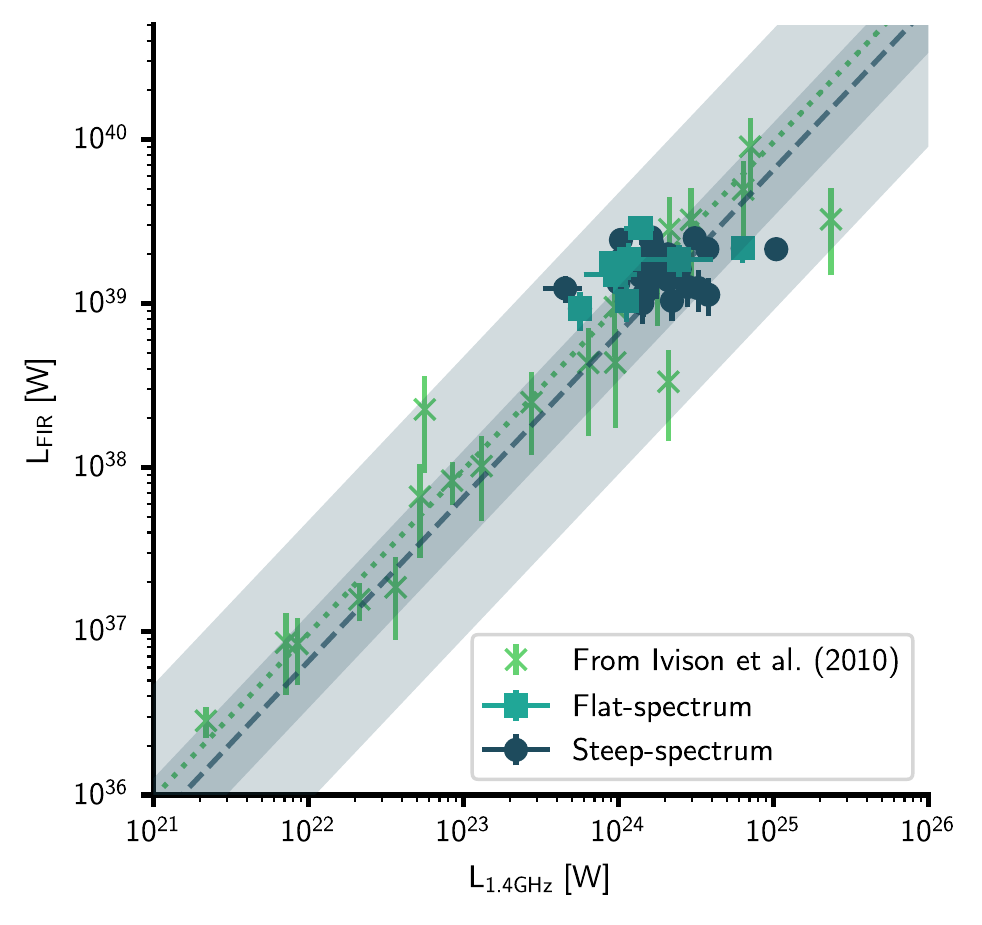}
            \caption{Far-infrared to radio correlation. Sources in our sample with extremely flat low-frequency radio spectral indices are plotted with light-coloured squares, and the positions of the `normal' sources are plotted with dark circles. The FIRC relation calculated from our data following Eq.~\ref{eq:firc} is plotted as a dark dashed line, while the shaded region shows the 1$\sigma$ and 3$\sigma$ variances, respectively. Data from \citet{2010MNRAS.402..245I} are plotted in green crosses for comparison, with the FIRC from that work plotted as a dotted  green line.}
            \label{fig:FIRC}
        \end{figure}

        The radio luminosity is calculated using the observed 1.4 GHz flux density, {\it K}-corrected according to the LOFAR catalogue redshift following $L_{\nu} = S_{\nu} \times 4\pi \times D_{l}^{2} \times (1+z)^{\alpha - 1}$.
        For each source, we used the value of the radio spectral index, $\alpha$, resulting from a power law fit to the high-frequency radio spectrum (324 MHz -- 1.4 GHz) of the form \smash{$S_{\nu} \propto \nu^{\alpha}$} (see Sect.~\ref{sec:radspec}). We plot our sample on the FIRC following \citet{2010MNRAS.402..245I}, and over-plot their results in green for comparison (Fig.~\ref{fig:FIRC}).

        We notice immediately that, despite differences in the location of our sources on the FIRC compared to \citet{2010MNRAS.402..245I} (likely due to our FIR estimation method and selection effects, as discussed below), none of our sources appear significantly below the FIRC, and therefore we may assume that our sources do not host radio-loud AGN, or at least that their spectra are dominated by contributions from star formation.
        Selecting only the brightest submillimetre sources from a flux-limited sample introduces selection effects such that we bias our sample towards the FIR-bright, radio-faint population; therefore, we do not attempt to comment on the distribution of our sources with regards to the normalisation of the FIRC and its dependence on other galaxy properties, such as redshift, temperature, and stellar mass, as extensively studied by, for example, \citet{2001ApJ...554..803Y}, \citet{2010MNRAS.402..245I}, \citet{2014MNRAS.445.2232S}, and \citet{2018MNRAS.480.5625R}.
        The depth of the radio surveys used in our sample also impacts the FIRC here: Many older studies are based on much shallower surveys, and the FIRC is plotted only for objects detected in both bands.
        With shallow survey flux limits, this biases the selection to only the brightest sources in both IR and radio.
        Here, selecting with S2CLS, we deliberately probed only the brightest, most star-forming IR galaxies, but with the depth of the VLA surveys we are able to detect the faint radio counterparts of these submillimetre-bright objects to lower flux limits than previous work. 
        These faint radio sources detected in our sample have the resulting effect that the distribution of our sources in Fig.~\ref{fig:FIRC} has a scatter that lies above the traditional FIRC.
        However, despite our sources filling a relatively small range of FIR luminosities, they span a wide range of radio luminosities, allowing us to study the variation in the radio properties of highly star-forming galaxies.
        Calculating upper limits on the radio luminosities of the two sources with no radio detections, as described in Sect.~\ref{subsec:nondetections}, we find that these sources also lie in the same region on the FIRC as the rest of our sample, suggesting that these are likely to be normal star-forming galaxies at very high redshift, as hypothesised.

        We calculated the FIRC parameter, $q_{\text{IR}}$, which parameterises the FIRC, as follows:

        \begin{equation}
        \label{eq:firc}
            q_{\text{IR}} = \text{log} \left(\frac{S_{{8 - 1000 \,\mu \text{m}}}/3.75 \times 10^{12}}{\text{W m}^{-2}}\right) - \text{log}\left(\frac{S_{1.4\, \text{GHz}    }}{\text{W m}^{-2} \text{Hz}^{-1}}\right)
        .\end{equation}{}
        
        We obtained a mean value of $q_{\text{IR}} = 2.24 \pm 0.29$ for the objects in our sample.
        This is consistent with the comparable value of $2.41 \pm 0.2$ obtained by \citet{2010MNRAS.402..245I}, despite our very simple method for estimating FIR luminosities and the selection effects discussed above.

        To explore the range of the radio properties of these sources, we look in more detail at the radio spectra in the following section.

    \section{Radio spectra}
    \label{sec:radspec}

\begin{figure*}
        \centering
            \includegraphics[width=0.8\textwidth]{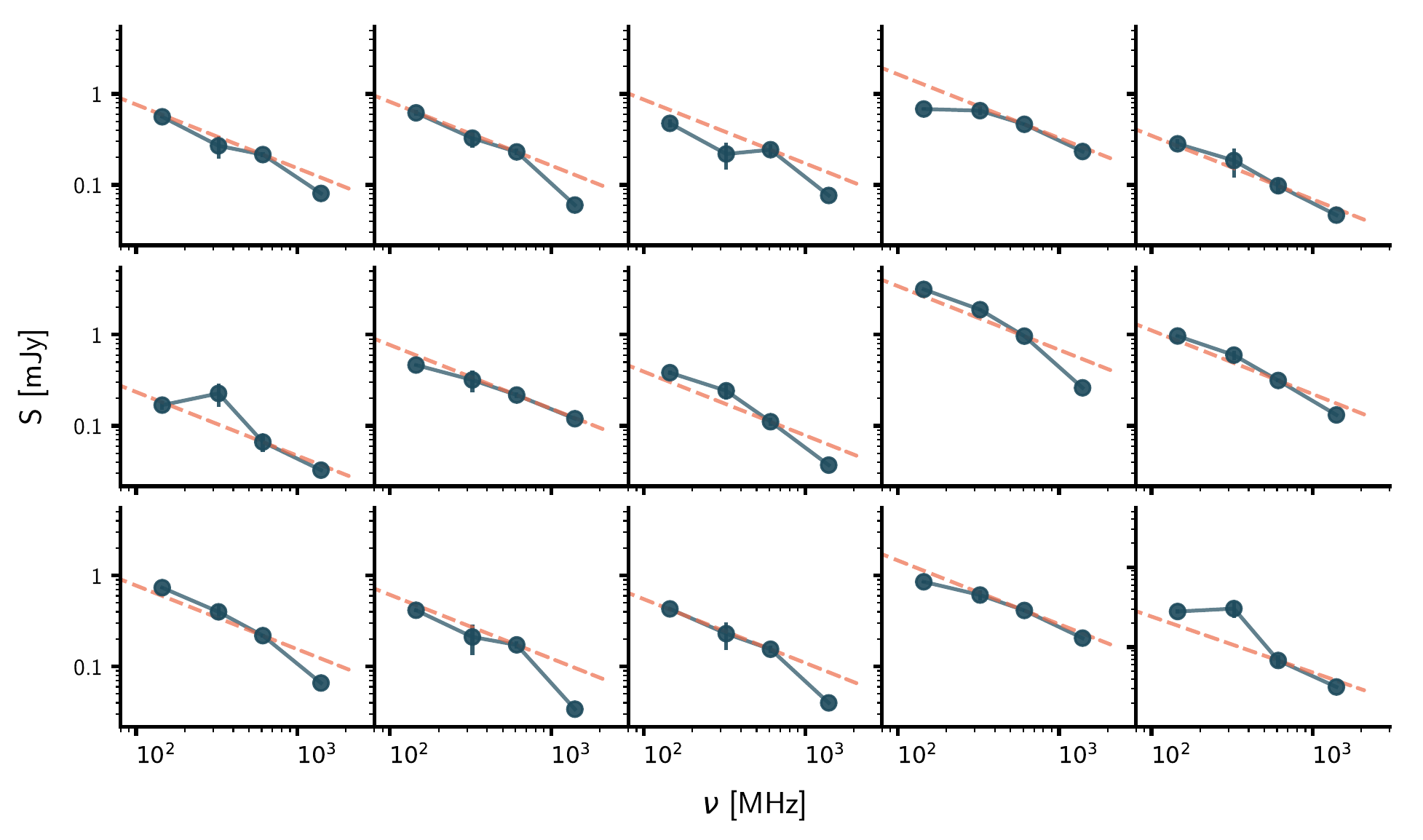}
            \caption{Examples of radio SEDs from our sample, with data at 150 MHz from LOFAR, 324 MHz and 1.4 GHz from the JVLA, and 610 MHz from the GMRT (see Sect.~\ref{subsec:additiona_data}). There is a range of luminosities and spectral shapes. Orange lines show a simple power law with spectral index $\alpha = -0.7$ normalised to the 610 MHz point as a reference and to demonstrate the spectral curvature present in many of these radio spectra. Error bars are shown but are largely within the size of the plotted points.}
            \label{fig:radio_SEDs}
            
        \end{figure*}

        The simplest approach to modelling the radio SED is to fit a power law, \smash{$S_{\nu} \propto \nu^{\alpha}$}, and allow the spectral index, $\alpha$, to vary.
        Fitting a power law across all available radio data for each source, we measured a median\footnote{Median errors calculated following \citet{2001ApJ...549....1G}.} spectral index across the whole sample of $-0.86 \pm 0.06$, consistent with values of $\alpha$ in previous studies \citep{1992ARA&A..30..575C, 2009MNRAS.392.1403S, 2013MNRAS.435..650M}.
        If we divide the SED into two sections, low frequency (150 MHz -- 324 MHz) and high frequency (324 MHz -- 1.4 GHz), we find median spectral indices of $-0.56 \pm 0.16$ at low frequency and $-0.97 \pm 0.15$ at high frequency.
        This choice of dividing frequency (324 MHz) is still firmly in the traditional low-frequency regime; however, if we were to split the sample at a higher frequency (e.g. 610 MHz), we would risk averaging over the part of the spectrum where we might see the effects of spectral flattening most prominently.
        In addition, our radio data span the range 150 MHz -- 1.4 GHz, and we did not include any higher frequency radio data in our analysis, so our use of the terms `low frequency' and `high frequency' are purely relative in this case.
        The lower average spectral index at low frequency reveals an average spectral flattening, but across the sample we see a range of different spectral shapes in the radio regime (Fig.~\ref{fig:radio_SEDs}).
        Extremely star-forming galaxies have been found to have steeper spectral indices in our high-frequency range than in previous studies \citep{2018MNRAS.474..779G}, and models by \citet{2010ApJ...717..196L} also predict steeper synchrotron spectra in high redshift `puffy' SMGs than in compact starbursts.

\begin{figure}
         \centering
         \includegraphics[width=\columnwidth]{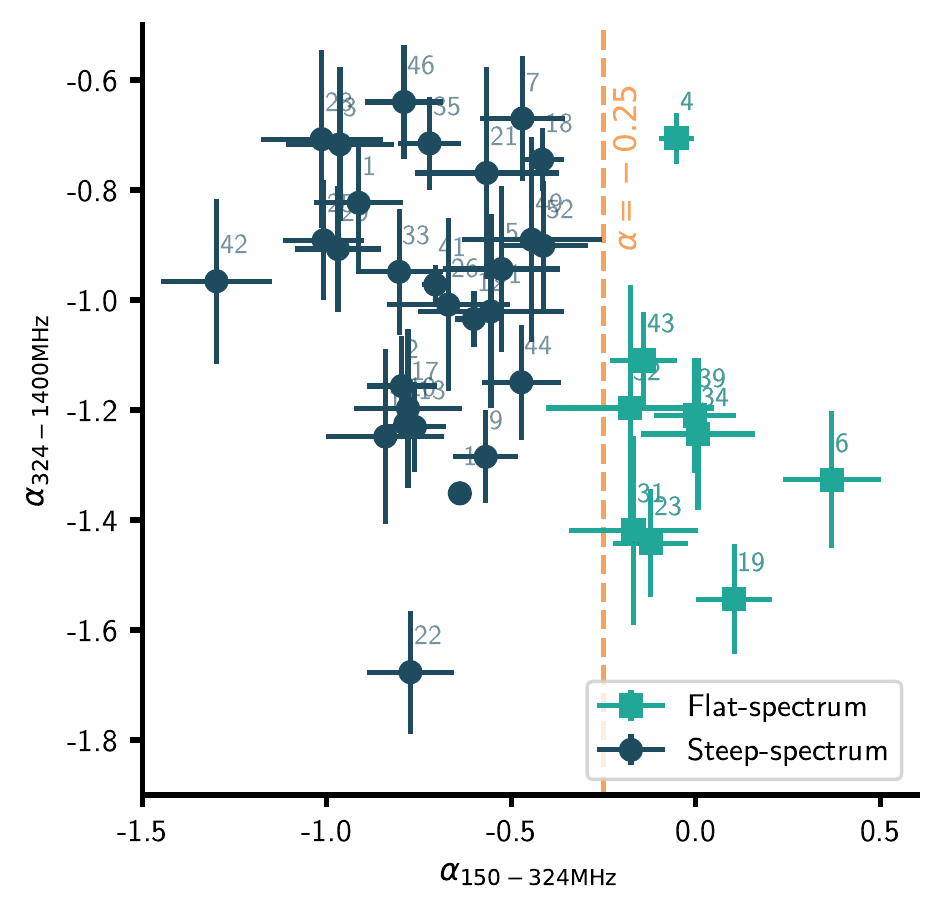}
         \caption{Radio colour-colour plot showing the low-frequency (150 MHz -- 324 MHz) spectral slope on the x-axis and the high-frequency (324  MHz-- 1.4 GHz) spectral slope on the y-axis. The dashed line indicates $\alpha_{low} = -0.25$, the cutoff above which we define sources to have extremely flat low-frequency spectra. Sources above this threshold are marked with squares in the plots throughout this paper. Sources are numbered with the IDs assigned in this study, as detailed in Table~\ref{table:sources}.}
         \label{fig:radio_slopes}
        \end{figure}

        This variety suggests that there is not one single set of physical conditions responsible for the shape of the radio spectrum, but that it is more likely that a variety of physical conditions contribute to the observed diversity of radio spectral shapes.
        With only four observed frequencies, we did not attempt to fit a more complex model but instead examined how the slope of the spectrum changes with frequency.
        We constructed a radio colour-colour plot using the observed-frame radio flux densities at 150 MHz, 324\,MHz, and 1.4\,GHz. 
        Defining a colour, equivalent to the local spectral index, as the relationship between the two flux densities,
        \begin{equation}
            \alpha = \frac{\log (S_{\nu_1} / S_{\nu_2})} {\log (\nu_1 / \nu_2)}
        ,\end{equation}
        we plot the low-frequency colour ($\alpha_{\rm low}, \:\nu_1 = 150 \, \rm MHz,\: \nu_2 = 324\, MHz$) against high-frequency colour ($\alpha_{\rm high},\: \nu_1 = 324 \,\rm MHz,\: \nu_2 = 1.4 \,GHz$) to obtain a diagnostic plot from which we may read a measure of spectral curvature (Fig.~\ref{fig:radio_slopes}).
        A non-negligible number of sources have significantly flatter low-frequency spectral slopes than expected from typical estimates of power-law-type radio spectra with $\alpha \simeq -0.8$.

\begin{figure}
            \centering
            \includegraphics[width=\columnwidth]{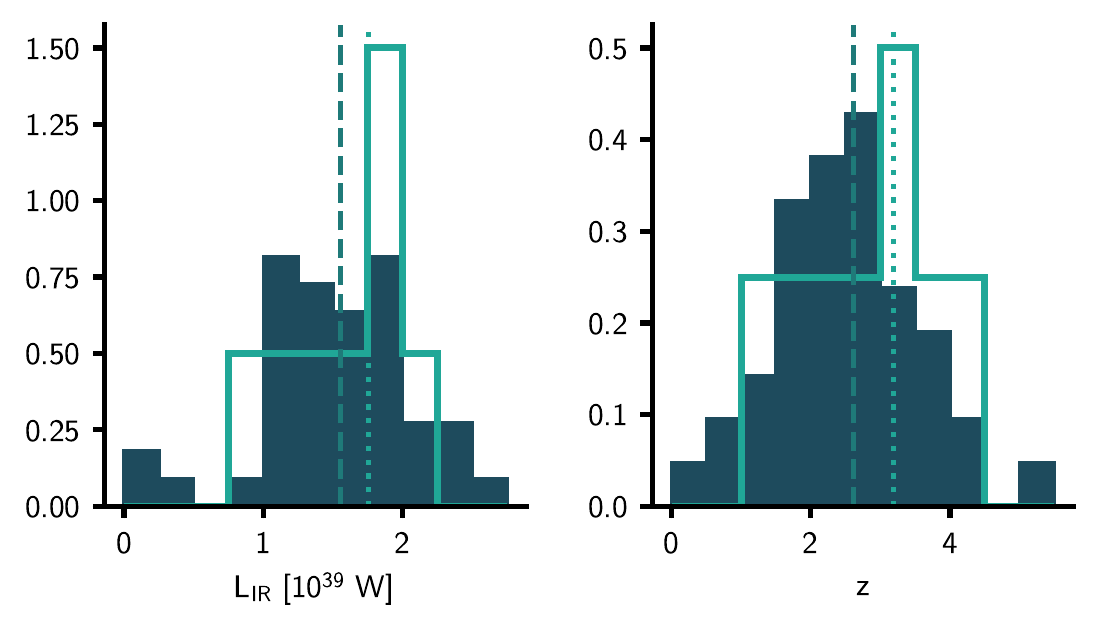}
            \caption{Normalised histograms showing the distribution of sources in IR luminosity and redshift, with the full sample shown in dark and the extremely flat-spectrum sources shown in a light outline. The median of each sample is also shown, with a dashed line showing the median of the full sample and a dotted line showing the median of the extremely flat-spectrum sample.}
            \label{fig:hists}
        \end{figure}

        We selected the nine sources with $\alpha_{\rm low} > -0.25$ (marked in Fig.~\ref{fig:radio_slopes} with a dashed red line) to investigate further and performed a number of comparisons to see whether this extremely flat $\alpha_{\rm low}$ sub-sample differs from the parent sample in any observational parameters.
        We note that conventionally a more conservative cut is made in spectral index to define flat-spectrum sources (typically $\alpha = -0.5$); however, we divided our sample as such so as to investigate the most extreme spectral flattening.
        Taking a more traditional division between flat and steep spectrum sources at $\alpha = -0.5$ does not, however, significantly affect our conclusions, and in fact a flat-spectrum sample with $\alpha > -0.5$ is comparable  to the extremely flat-spectrum  $\alpha > -0.25$ sample in terms of luminosity and redshift distribution (as in Fig.~\ref{fig:hists}) as well as the other properties explored in this study.

        As an initial check, we confirmed that the distribution of these sources is similar to that of the parent sample in both luminosity and redshift (Fig.~\ref{fig:hists}).
        The extremely flat-spectrum sample is statistically indistinguishably distributed from the parent sample in both, with a Kolmogarov-Smirnov two-sample test (KS test) giving $p$-values of $p = 0.98$ and $p = 0.61$ for IR luminosity and redshift, respectively.
        
\begin{figure}
            \centering
            \includegraphics[width=\columnwidth]{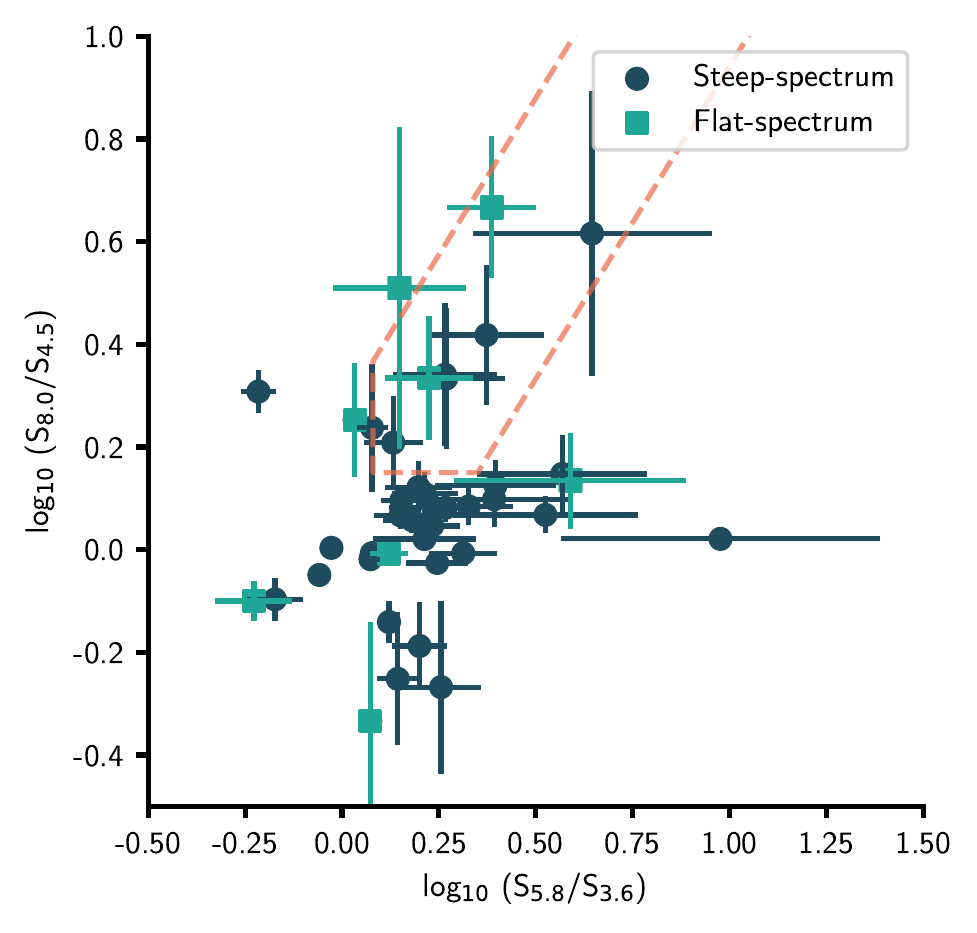}
            \caption{IRAC colour-colour plot, with the AGN `wedge' from \citet{2012ApJ...748..142D} shown with a dashed line. Light squares show the extremely flat-spectrum sources. }
            \label{fig:irac_colours}
        \end{figure}

        Next we investigated the MIR colours of these sources.
        If the difference in radio spectral shape is correlated with a difference in dust temperatures, we might expect to see this sub-sample in a distinct region of an MIR colour-colour diagram.
        Mid-infrared colour photometry can be used as a diagnostic to identify the presence of AGN in star-forming galaxies \citep{2000A&A...359..887L}.
        The MIR spectrum of a star-forming galaxy is typically driven by emission from warm dust ($T \sim 25$ -- $50$\,K) heated by H~{\sc ii} regions associated with recent star formation, as well as photodissociation regions giving rise to sharply peaked emission features from polycyclic aromatic hydrocarbons (PAHs). 
        A luminous AGN will contribute significantly to the short-wavelength emission of the MIR spectrum via dust in the AGN torus being heated to much higher temperatures of up to $T \sim 1500$\,K.
        Colour-colour plots constructed from {\it Spitzer} Infrared Array Camera (IRAC) photometry can be used to distinguish whether the dominant contribution is from AGN, in which case the MIR colours will be very red \citep{2004ApJS..154..166L}. We plot the positions of our sources in MIR colour-colour plots using the 3.6, 4.5, 5.8, and 8 $\mu$m\xspace IRAC flux densities (Fig.~\ref{fig:irac_colours}).

        We followed \citet{2012ApJ...748..142D} and over-plotted the region defined therein to identify galaxies hosting AGN using their IR colours.
        Our main interest in investigating the MIR colours is to determine whether the difference in radio spectral slope could result from AGN contributions.
        We have already established that there is no clear signature of radio-loud AGN in any sources; however, by inspecting the MIR colours we can probe their dust temperature distributions, which could reveal AGN activity with low radio luminosities that could affect the shape of the radio spectrum.
        A number of our sources lie within this region, suggesting that these sources may contain AGN.
        There are also a number of other mechanisms that could affect the MIR properties in addition to AGN activity -- it should be noted that we cannot disentangle any differences in MIR emission due to the relative timescales of star-formation and AGN activity, nor that of the merger stage.
        The range of redshifts across the sample will also result in a spread of MIR colours due to the location of PAH features in the observed frame.
        Nevertheless, there is no clear distributional difference between the extremely flat-spectrum sample and the rest of the sample --  a KS test giving $p = 0.4$ in the $S_{5.8}/S_{\rm 3.6}$ colour and $p = 0.95$ in the $S_{8.0}/S_{\rm 4.5}$ colour -- with neither sample clearly occupying a distinct region in the colour-colour plot, suggesting that the cause of this radio spectral flattening must be something other than simply AGN contributions to the spectrum. As the sample is of only a small number of sources, it is difficult to make a statistically robust statement on their distribution; however, there is nothing to suggest that the extremely flat-spectrum sources are special in any way as regards their IR properties.

As there are no discernible statistically significant differences between the extremely flat $\alpha_{\text{low}}$ sources and the rest of the sample in our observable properties, we propose that this difference in radio spectral shape may be related to properties of the galaxies that we are unable to detect with our unresolved, galaxy-averaged observations.

    \subsection{Low-frequency flattening of radio spectra}
    \label{subsec:flat_spectra}
        There are several possible scenarios in which the low-frequency radio spectral slope is flattened relative to an $\alpha =-0.7$ power law spectrum resulting from the conditions of the ISM, as observed in the nine sources described in Sect.~\ref{sec:radspec}.
        Synchrotron self-absorption can play a role in flattening the low-frequency spectrum; however, as the galaxies in our sample do not have AGN-dominated spectra, this is unlikely to make a significant contribution.
        In the case of low-luminosity AGN contribution, the level of flattening observed in our extremely flat-spectrum sources could only be obtained with a combination of the maximum possible AGN fractional contribution and self-absorption tuned to occur exactly at our breaking frequency of 324 MHz.
        This would be an unlikely coincidence, and as such we conclude that it is unlikely for AGN contributions alone to cause the observed flattening.
        A more likely possibility is that, if the source of synchrotron emission is embedded in an ISM that is sufficiently dense and clumpy, free-free absorption begins to have a significant effect as we observe the lower-frequency spectrum.
        Nearby star-forming galaxies have been observed to have significant free-free absorbed spectra and clumpy star-forming regions \citep[e.g.][]{2009AJ....137..537L, 2014AJ....147....5R} but it is only now with the deep low-frequency radio data that have become available with LOFAR that we are able to investigate this at high redshift.

        To model free-free absorbed spectra, we assumed typical properties of the ISM of SMGs and computed the effect this has on an $\alpha =-0.7$ power law spectrum at the median (estimated) redshift $z = 2.6$ of the sample.
        We calculated the optical depth \smash{$\tau = \int \kappa dl$} at the rest-frame frequencies that correspond to our observed frequencies for a galaxy at the median redshift of the sample ($z=2.6$), where $\kappa$ is the free-free absorption coefficient as defined by \citet{1992ARA&A..30..575C}:

        \begin{equation}
            \left(\frac{\kappa}{\text{pc}^{-1}}\right) = 3.3 \times 10^{-7} \left(\frac{n_e}{\text{cm}^{-3}}\right)^2 \left(\frac{T_e}{10^4 \text{K}}\right)^{-1.35} \left(\frac{\nu}{\text{GHz}}\right)^{-2.1} 
        ,\end{equation}
        where $n_e$ is electron density and $T_e$ electron temperature.
        Estimates of electron densities in star-forming galaxies from emission line diagnostics are poorly constrained and range from $10 \text{ to } 400 \text{ cm}^{-3}$ \citep{2013ApJ...776...65P, 2013ApJ...776...38F, 2014ApJ...785..153M, 2017MNRAS.465.3220K}.
        We assumed an electron density of $n_e = 50\ \text{cm}^{-3}$, which is on the low end of star-forming electron density estimates -- however, as the length scale of relevance is inversely proportional to $n_e^{2}$ for an equivalent level of absorption, if we were to assume a higher value of $n_e$ we would find length scales that are smaller by a factor of $\Delta n_e^{2}$.
        This choice then gives us an approximate upper limit for the length scales required for free-free absorption to flatten the spectrum at low frequency, as observed.
        We assumed an electron temperature of $T_{e} = 10^4 \text{ K}$ \citep{1980A&AS...40..379D}.
        Of course, the properties of the ISM in an individual galaxy will vary in density and temperature across the galaxy's extent as well as across our sample, which covers a range of redshifts; we expect to see variation in the nature of the ISM.
        This simple model of the effects of free-free absorption assumes a single slab of absorbing material; due to the nature of our observations, it would be unrealistic to attempt to model any more realistic distributions of dense gas in the ISM.
        However, more complex distributions of gas mixed with the radio-emitting plasma in simulations of free-free absorption in radio spectra \citep[e.g.][]{2018MNRAS.475.3493B, 2014A&A...566A..15V} produce very similar results overall.

        \begin{figure*}
            \centering
            \includegraphics[width=0.8\textwidth]{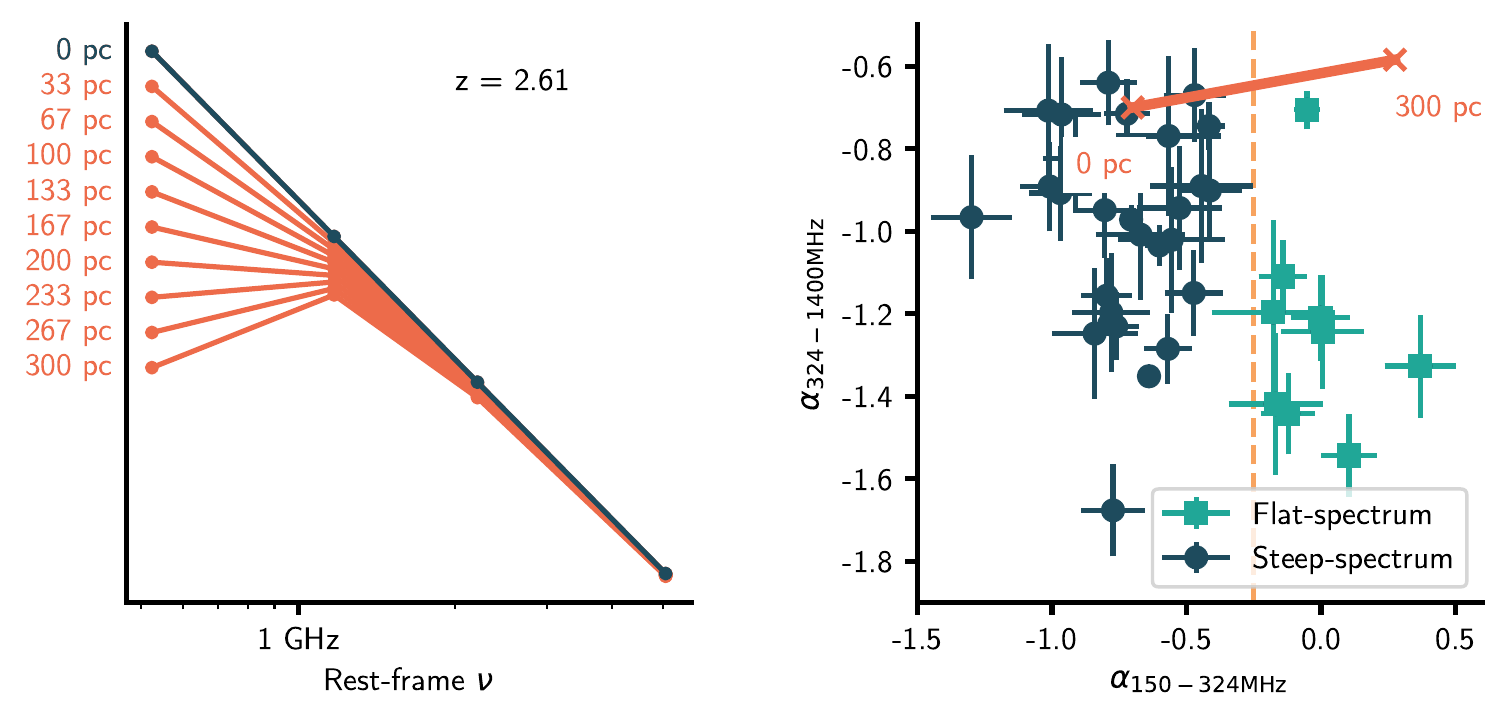}
            \caption{Our simple model of free-free absorption and its effect on spectral indices. \textit{Left:} Effect of free-free absorption on a simple power law spectrum with radio spectral index $\alpha = -0.7$, with increasing lengths of absorbing column, for our observing frequencies in the observed frame. We assume an electron density $n_e = 50 \text{ cm}^{-3}$ and temperature $T = 10^4 \text{ K}$. \textit{Right:} As Fig.~\ref{fig:radio_slopes}, but with the effect of free-free absorption on a simple power law spectrum plotted in orange, with increasing lengths of absorbing column from zero (left) to 300 pc (right).}
            \label{fig:absorbed_spec}
        \end{figure*}

        Figure~\ref{fig:absorbed_spec} shows the effect of free-free absorption on such a spectrum, over absorbing columns of between zero and 300 pc.
        In this model, the level of spectral flattening that we would be able to observe occurs at rest-frame frequencies of $< 1$ GHz, which corresponds to observed-frame frequencies of $\lesssim 280$ MHz for the median redshift, $z = 2.6$, of our sample.
        This in part motivated our choice of cutoff frequency to measure the spectral slope as measuring at 610 MHz would average out the flattening we see most strongly at these lower frequencies.
        We return to the radio colour-colour plot in Fig.~\ref{fig:radio_slopes} and plot the colours of this model spectrum over these increasing absorbing columns.
        This simple model demonstrates that, with our assumed electron densities and temperature, an absorbing column of several hundred parsecs in size would be sufficient to flatten the low-frequency radio spectral index to the degree that we observe in several sources.
        As noted above, if we increase the electron density in our model, this would be consistent with smaller-scale absorbing columns (approximately tens of parsecs).
        
        This spatial scale of dense, star-forming clumps is consistent with the highest-resolution observations of SMGs \citep{2010Natur.464..733S, 2015ApJ...810..133I, 2019NatAs...3.1114D}. 
        A number of studies imaging SMGs at the highest resolutions have found evidence of clumpy substructure at spatial scales of several hundred parsecs \citep[e.g.][]{2016ApJ...829L..10I, 2018Natur.560..613T}; however, for the most part it is only possible to resolve down to scales of approximately tens of kiloparsecs in the handful of submillimetre sources that are strongly gravitationally lensed and observed with interferometric instruments.
        One such strongly lensed submillimetre source that has been well studied, SDP.81, displays a spatially non-uniform dust continuum, with several bright $100 - 500$ pc scale clumps \citep{2015MNRAS.451L..40R, 2015PASJ...67...93H, 2015ApJ...806L..17S} and `knots' \citep{2015PASJ...67...72T} on $< 100$ pc scales.
        The `Cosmic Snake' is another -- with magnification affording resolved scales as small as 30 pc; \citet{2019NatAs...3.1114D} find several molecular clouds on scales between 30 -- 210 pc.
        
        These scales correspond approximately to the Jeans length for the gas densities typical of star-forming galaxies \citep{2015MNRAS.451.1284S, 2015PASJ...67...93H}, with giant molecular gas clouds collapsing as a result of Toomre instabilities \citep{1964ApJ...139.1217T}. However, we caution the reader that there has been debate as to whether observed `clumpy' substructure is biased by effects relating to interferometric observations.
        The case for ${\sim} 100\  $pc scale substructure in the `Cosmic Eyelash' \citep{2010Natur.464..733S} has recently been challenged, with work showing that the inferred structure may be due to filtering and resolution effects amplifying spurious features in low S/N interferometric images \citep{2018NatAs...2...76C,2018ApJ...859...12G, 2020MNRAS.495L...1I}.
        
        Our extreme low-frequency spectral flattened sample with $\alpha > -0.25$ consists of only nine of the 42 sources in the whole sample, implying that if our assumption that this spectral flattening is caused by free-free absorption is correct, the majority of the population of submillimetre-bright sources might not be expected to show strong evidence of dense, clumpy star-forming regions on these scales.
        The effect of absorption on our unresolved galaxy-averaged observations must depend on the fraction of low-frequency radio emission that is embedded within and absorbed by high density gas; if a sufficient fraction is able to escape, we will see this flattening to a lesser degree.
        Thus there are several explanations for the lack of free-free absorption signatures in the majority of our galaxy sample.
        It could be due to an intrinsically smoother, more diffuse distribution of star-forming material \citep[as suggested in e.g.][]{2016ApJ...833..103H, 2020MNRAS.495L...1I}.
        There is likely also a dependence on age -- in younger star-forming regions, the radio emission is more likely to still be contained within a dense gaseous structure -- and geometry may also play a role.
        Effects of environment and the merger stage may affect the distribution of gas in our galaxies, and we spanned a large range in redshift over which we would expect the nature of the ISM to evolve.
        Additionally, our estimated ${\sim} 100$ pc length scale is based on a conservatively low estimate of electron density, as previously discussed; higher assumed electron densities would lead to clumps on scales below the resolution limits of observations of even the most highly magnified lensed sources.
        
        Due to the serendipitous nature of locating gravitationally lensed galaxies, there are very few sources in which observations of sufficiently high resolution can be made to detect substructure on the ${\sim} 100$ pc scales implied by our calculation.
        Since only ${\sim} 20$ per cent of sources in our sample display this low-frequency radio spectral flattening, a much larger sample of hundreds of galaxies observed at sub-kiloparsec scales would be required to detect many sources with this substructure if our assumptions are correct that this observed radio spectral flattening is due to free-free absorption, and assuming that structure does exist on scales large enough to be detected with current instrumental capabilities.

\section{Conclusions}

Taking advantage of new, deep LOFAR images, we investigated the low-frequency radio spectra of a sample of highly star-forming galaxies selected at 850 $\mu$m\xspace from the S2CLS.
Our conclusions are as follows:
   \begin{enumerate}
      \item 
        We find that this sample of SMGs displays a range of radio luminosities and spectral shapes despite being selected at a very narrow range of submillimetre fluxes, implying that the radio properties of this sample do not follow a tight correlation with the star-formation properties we might infer from submillimetre observations alone.
      \item 
        We find evidence of radio spectral flattening at low frequencies ($\alpha_{\rm low} = -0.47 \pm 0.16$ on average, and nine sources with $\alpha_{\rm low} < -0.25$).
        These flat-spectrum sources are indistinguishable from the full sample in their distributions of redshift and IR luminosity, as well as in their IR colours.
        In the absence of any clear observational differences between sources with flat low-frequency spectra and the rest of the sample, we infer that this must be due to underlying properties of the galaxies that we cannot observe in our unresolved imaging.
        We suggest that this radio spectral flattening may be due to free-free absorption arising from non-uniform, clumpy distributions of ionised gas in which star formation is embedded.
        Taking typical values of electron density and temperature, we estimate that  clumps must be of the order of a few hundred parsecs in size (comparable to substructure that has been observed in strongly lensed submillimetre sources) to account for the observed radio spectral curvature.
        This presents an additional piece of evidence in favour of clumpy star formation in high redshift galaxies that does not depend on the angular resolution of morphological imaging but can be detected in the galaxy-averaged properties in the radio SED.
        Due to the serendipitous nature of imaging lensed sources, there are few observations reaching high enough resolution to detect structure on this scale, and so larger samples of high-resolution images (e.g. using submillimetre interferometric instruments such as ALMA) would be required to test whether the proportion of galaxies in our sample that exhibit this radio spectral flattening (${\sim} 20$ per cent) is typical of the submillimetre population in general.
      \item 
        In addition to the range of radio spectra observed in this sample, we also find two bright submillimetre sources ($> 7 \sigma$ detections) that are undetected at all other wavelengths, from optical through to radio. 
        We speculate that, due to sampling the peak of the thermal dust emission spectrum at 850 $\mu$m\xspace, these objects are located at high redshift ($z > 4$) and are too faint due to cosmological dimming at all other observed wavelengths.
        We propose them as interesting candidates for (sub)millimetre interferometric follow-up to determine their redshifts with certainty.
    
   \end{enumerate}
Finding this variety of spectral shapes and evidence of free-free absorption in spectral flattening at low frequencies is consistent with high-resolution observations of clumpy star-forming regions in submillimetre galaxies.
A larger sample of submillimetre galaxies with low-frequency radio observations and high-resolution interferometric follow-up would be beneficial to further investigating the nature of this spectral flattening.

\begin{acknowledgements}
JR, MJH, PNB, IM and RK acknowledge support from the UK Science and Technology Facilities Council [STFC: ST/N504105/1, ST/R000905/1, ST/R000972/1, ST/R505146/1 and ST/R504737/1]. JEG is supported by the Royal Society through a University Research Fellowship.
KJD and HR acknowledge support from the ERC Advanced Investigator programme NewClusters 321271. MJJ acknowledges support from the UK Science and Technology Facilities Council [ST/N000919/1] and the Oxford Hintze Centre for Astrophysical Surveys which is funded through generous support from the Hintze Family Charitable Foundation. MB acknowledges support from INAF under PRIN SKA/CTA FORECaST and from the Ministero degli Affari Esteri della Cooperazione Internazionale - Direzione Generale per la Promozione del Sistema Paese Progetto di Grande Rilevanza ZA18GR02.IP acknowledges support from INAF under the SKA/CTA PRIN “FORECaST” and the PRIN MAIN STREAM “SAuROS” projects.

LOFAR, the Low Frequency Array designed and constructed by ASTRON, has
facilities in several countries, which are owned by various parties
(each with their own funding sources), and are collectively operated
by the International LOFAR Telescope (ILT) foundation under a joint
scientific policy. The ILT resources have benefited from the
following recent major funding sources: CNRS-INSU, Observatoire de
Paris and Universit\'e d'Orl\'eans, France; BMBF, MIWF-NRW, MPG, Germany;
Science Foundation Ireland (SFI), Department of Business, Enterprise
and Innovation (DBEI), Ireland; NWO, The Netherlands; the Science and
Technology Facilities Council, UK; Ministry of Science and Higher
Education, Poland.

Part of this work was carried out on the Dutch national
e-infrastructure with the support of the SURF Cooperative through
grant e-infra 160022 \& 160152. The LOFAR software and dedicated
reduction packages on \url{https://github.com/apmechev/GRID_LRT} were
deployed on the e-infrastructure by the LOFAR e-infragroup, consisting
of J.\ B.\ R.\ Oonk (ASTRON \& Leiden Observatory), A.\ P.\ Mechev (Leiden
Observatory) and T. Shimwell (ASTRON) with support from N.\ Danezi
(SURFsara) and C.\ Schrijvers (SURFsara). This research has made use of the University
of Hertfordshire high-performance computing facility
(\url{https://uhhpc.herts.ac.uk/}) and the LOFAR-UK compute facility,
located at the University of Hertfordshire and supported by STFC
[ST/P000096/1]. The J\"ulich LOFAR Long Term Archive and the German
LOFAR network are both coordinated and operated by the J\"ulich
Supercomputing Centre (JSC), and computing resources on the
supercomputer JUWELS at JSC were provided by the Gauss Centre for
supercomputing e.V. (grant CHTB00) through the John von Neumann
Institute for Computing (NIC).

The James Clerk Maxwell Telescope is operated by the East Asian Observatory on behalf of The National Astronomical Observatory of Japan; Academia Sinica Institute  of  Astronomy  and  Astrophysics;  the  Korea  Astronomy and Space Science Institute; the Operation, Maintenance  and  Upgrading  Fund  for  Astronomical  Telescopes and Facility Instruments, budgeted from the Ministry of Finance  (MOF)  of  China  and  administrated  by  the  Chinese Academy  of  Sciences  (CAS),  as  well  as  the  National  Key R\&D Program of China (No. 2017YFA0402700). Additional funding support is provided by the Science and Technology Facilities  Council  of  the  United  Kingdom  and  participating  universities  in  the  United  Kingdom  and  Canada.

The authors wish to recognize the very significant role that the summit of Maunakea has within the Indigenous Hawaiian (Kānaka Maoli) community, and acknowledge the continued use of indigenous land for telescope facilities.

We made use of SciPy \citep{jones_scipy_2001}, NumPy \citep{van2011numpy} and matplotlib, a Python library for publication-quality  graphics  \citep{Hunter:2007},  as  well  as  Astropy, a community-developed core Python package for Astronomy \citep{2013A&A...558A..33A}.
\end{acknowledgements}

%
%
\bibliographystyle{aa} 
\bibliography{references} 
\end{document}